\documentclass[12pt,oneside,english,12pt, oneside, reqno]{amsart}
\usepackage[T1]{fontenc}
\usepackage[latin9]{inputenc}
\usepackage{geometry}
\usepackage{amstext}
\usepackage{amsthm}
\usepackage{amssymb}
\usepackage{stmaryrd}
\usepackage{setspace}
\usepackage[authoryear,round,comma]{natbib}
\makeatletter
\theoremstyle{plain}
\newtheorem{prop}{\protect\propositionname}
\theoremstyle{definition}
\newtheorem{defn}{\protect\definitionname}
\theoremstyle{remark}
\newtheorem{claim}{\protect\claimname}

\usepackage{lmodern}
\usepackage[T1]{fontenc}
\usepackage{braket}
\usepackage{hyperref}
\hypersetup{
colorlinks=false,
citecolor=green
}

\makeatother

\usepackage{babel}
\providecommand{\claimname}{Claim}
\providecommand{\definitionname}{Definition}
\providecommand{\propositionname}{Proposition}

\begin{document}
\title{Identifying Assumptions and Research Dynamics}
\author{Andrew Ellis and Ran Spiegler}
\date{14 January 2025}
\thanks{Ellis: LSE, a.ellis@lse.ac.uk. Spiegler: Tel Aviv University and UCL,
rani@tauex.tau.ac.il. Spiegler acknowledges financial support from
ISF grant no. 320/21. We thank Daron Acemoglu, Tim Christensen, In-Koo
Cho, Martin Cripps, Tuval Danenberg, Andrew Gelman, Duarte Goncalves,
Charles Manski, Ignacio Esponda, Jesse Shapiro, Kate Smith, Zihan
Jia, and audiences at Bar-Ilan University, BRIC, BU, NYU, Maryland,
a Stony Brook workshop on bounded rationality and learning, and an
LSE/UCL theory conference for helpful comments.}
\begin{abstract}
A representative researcher has repeated opportunities for empirical
research. To process findings, she must impose an \textquotedblleft identifying
assumption.\textquotedblright{} She conducts research when the assumption
is sufficiently plausible (taking into account both current beliefs
and the quality of the opportunity), and updates beliefs as if the
assumption were perfectly valid. We study the dynamics of this learning
process. While the rate of research cannot always increase over time,
research slowdown is possible. We characterize environments in which
the rate is constant. Long-run beliefs can exhibit history-dependence
and ``false certitude.'' We apply the model to stylized examples
of empirical methodologies: experiments, various causal-inference
techniques, and \textquotedblleft calibration.\textquotedblright{}
\end{abstract}

\maketitle

\section{Introduction}

\raggedbottom 

How should scientific research be conducted? The standard model of
rational behavior assumes that agents undergo Bayesian learning (following
\citet{deFinetti1974a} and \citet{Savage1954}). Bayesianism is thus
a natural benchmark if one thinks of scientific researchers as rational.
A Bayesian researcher holds a probabilistic prior belief regarding
a research question, accumulates evidence (such as controlled experiments
or observational data), and updates her prior in light of that evidence
via Bayes\textquoteright{} rule. Bayesian researchers would eventually
learn the truth (or at least some part of it), provided that the evidence
is informative and their prior does not rule it out.\footnote{For philosophy-of-science discussions of Bayesianism as a normative
theory for scientific learning, see \citet{howson2006scientific}
and \citet{GelmanShalizi2013}.}

Empirical research in economics often instead interprets evidence
through the lens of \textit{identifying assumptions}. These are explicit
hypotheses regarding the relationship between the evidence and the
research question that enable researchers to draw clear-cut conclusions
from their observations. For instance, consider learning about the
causal effect of class size on students' test performance. To draw
causal inferences from an observational dataset that links these two
variables, researchers need to form beliefs about how students were
assigned to classes. A typical identifying assumption in this context
is that the assignment was random, which enables a sharp causal interpretation
of students' performance differences across class sizes.

Researchers do not mindlessly impose these assumptions. They debate
over their plausibility for the dataset in question, and adopt them
only when deemed sufficiently plausible. This judgment may evolve
over time. For example, as researchers accumulate evidence about the
effect of class size on student performance, they develop more precise
beliefs about this effect, which may lead them to become more (or
less) stringent in admitting an identifying assumption that is only
approximately valid. The ongoing, case-by-case evaluation of an assumption's
plausibility distinguishes this assumption-based learning process
from Bayesianism. There, assumptions may inform the prior belief but
then cease to play any subsequent role.

Assumption-based learning circumvents some practical difficulties
with implementing the strict Bayesian recipe. Identifying assumptions
reduce the dimensionality of the uncertainty that the research community
needs to process and communicate. They are also typically simple to
describe. These features simplify the process of belief revision.
For example, if the research community accepts the random-assignment
assumption in the example, then its subjective beliefs about the assignment
process do not affect the updating process. Identifying assumptions
also facilitate reaching a long-run consensus in the research community,
since according to them, the evidence alone determines the long-run
answer to the research question.

Despite these advantages, if researchers make assumptions for practical
convenience but later take them too seriously, then they effectively
engage in misspecified learning. As a result, they may come to hold
incorrect and (by the nature of identifying assumptions) strongly
held beliefs. That is, the process may lead to what \citet{Manski_2020}
called ``incredible certitude.''

In this paper, we model a research process that treats identifying
assumptions as necessary for conducting research but that otherwise
conforms to Bayesian learning. A representative researcher, a stand-in
for the relevant research community, wants to determine some of (or
all of) the values of a collection of fixed parameters. (e.g., the
effect of class size on students' test performance). She faces a sequence
of research designs of random quality, given by \textit{i.i.d.} context
parameters (e.g., the extent to which assignment of students to classes
in a dataset is random). Both fixed and context parameters directly
affect the data-generating process of the study, if it is carried
out.

The community accepts the findings only when they are conducted under
a sufficiently plausible identifying assumption (e.g., that student
assignment is perfectly random). The assumption fixes the value of
the context parameters in such a way that if the study were independently
repeated many times, the results would produce a definitive answer
to the research question. If the assumption is not sufficiently plausible,
then the study is ignored (or not carried out in the first place),
and the researcher waits for the next opportunity. When the study
is carried out, the identifying assumption informs how its results
are incorporated into updated beliefs.

The community evaluates an assumption's plausibility by comparing
its beliefs regarding the distribution of variables under the actual
and assumed values of the context parameters. If the former distribution
differs too much from the latter, the researcher deems the assumption
implausible and passes over the opportunity to conduct research. If
the difference is small, she deems the assumption plausible, conducts
the study, observes its result as determined by the true data-generating
process, and updates her beliefs as if the assumption held \textit{exactly}.
We quantify the difference using an \textit{f-divergence} \citep{csiszar1967information},
which measures the expectation of a convex function of the likelihood
ratios induced by the two distributions. The ubiquitous Kullback-Leibler
(KL) divergence is a special case, which we employ in our applications.
Note that the plausibility judgment depends not only on the specific
details of the research opportunity but also on the community's beliefs
over the fixed parameters. As a result, whether or not the research
is conducted or processed depends on past decisions.

We study the dynamics and long-run behavior of this assumption-based
learning process. Our focus is on how the propensity to conduct research
(via the imposition of an identifying assumption) changes as the research
community\textquoteright s beliefs evolve over time. We show that
this propensity cannot always increase over time. In other words,
the research community cannot consistently lower its standards for
accepting research as time goes by. It may, however, continually raise
these standards, leading to a slowdown in the rate of research but
an increase in its quality. Intuitively, as the researcher's belief
exhibits greater confidence, she becomes more sensitive to the assumption's
rough edges and therefore more reluctant to impose it.

We also provide a sufficient condition for a time-invariant propensity
to conduct research. Essentially, the condition states that if an
observable variable is correlated with context parameters given the
other observables, then it must be independent of the fixed parameters
of interest (and other fixed parameters that the identifying assumption
renders observationally relevant). Thus, our condition demands separation
between the observable effects of fixed parameters of interest and
the context parameters. This result employs tools from the literature
on graphical probabilistic models (e.g., \citet{pearl2009}).

We then ask what the researcher eventually comes to believe. We define
a stable belief to be one that the updating process converges to with
positive probability. Stable beliefs concentrate on parameter values
for which observables' distribution conditional on the assumption
is closest (in the sense of KL divergence) to their empirical distribution
given the true value and the contexts in which research is conducted.
In turn, these contexts are determined by the stable belief itself.
This two-way relation makes stable beliefs an equilibrium object.
For most models of interest, stable beliefs almost always assign probability
one to a wrong value of the fixed parameters of interest. This result
resonates with the Manski's above-mentioned incredible-certitude critique
of scientific learning based on strong identifying assumptions.

We demonstrate the model's scope with stylized examples of familiar
empirical methodologies. One example considers experimental research
contaminated by interference (the identifying assumption rules out
the interference). We show that the propensity to conduct the experiment
decreases over time. Another example examines causal inference contaminated
by confounding effects (the identifying assumption is that no such
confounding exists). A variation on this example addresses instrumental-variable
designs (the identifying assumption is that the instrument is independent
of a latent confounder). Both variants satisfy our sufficient condition
for time-invariant propensity to conduct research. All three examples
satisfy the condition that leads to incredible certitude in the long-run.

Later in the paper, we present two examples that expand the notion
of identifying assumptions to cover fixed parameters, and allow the
researcher to choose from a set of candidate identifying assumptions.
First, we consider a researcher who tries to identify two fixed parameters
from noisy observations of their sum, but can only do so in piecemeal
fashion. She thus employs an identification strategy reminiscent of
the ``calibration'' method in quantitative macroeconomics. In the
long run, this researcher correctly learns the parameters' sum, but
she becomes perfectly confident of an almost surely wrong estimate
of each individual parameter. Second, we present a stylized model
of inference from selective samples, where the researcher wishes to
learn the returns from an activity. She considers two alternative
identifying assumptions: agents' selection into this activity is purely
random, or selection is systematically related to observables that
do not directly affect returns. The latter is a structural identifying
assumption that captures in stylized form the method of Heckman selection
\citep{heckman1979}. Learning dynamics exhibit history-dependence:
using an identification method changes beliefs in a way that can reinforce
its use.

\subsection*{Related literature}

Our paper continues a recent literature on Bayesian learning under
misspecified prior beliefs (\citet{espondaPouzo2016,fudenbergEtAl2017,frickEtAl2022,heidhuesEtAl2021,bohrenHauser2021,espondaEtAl2021}).\footnote{For a review of alternatives to Bayesian updating, see \citet{Ortoleva2024}.}
One strand in this literature (e.g., \citet{ChoKasa2015}, \citet{Ba2024})
incorporates continual model selection and misspecification tests
into the learning process. Our paper departs from the literature in
several respects. First, the identification motive for adopting a
misspecified model and the belief-based plausibility criterion that
governs it are novel. Second, the rate of learning in our model is
endogenous and can vary over time. In general, our economics-of-science
angle is new to the misspecified-learning literature.

The econometrics literature contains methodological discussions of
the role of identifying assumptions (\citet{rothenberg1971}, \citet{manski2007},
\citet{lewbel2019}). However, we are unaware of earlier discussions
of how identification methods can be reconciled with the Bayesian
approach. As we saw, Manski himself is a critic of using strong identifying
assumptions. Our paper can be viewed as a\textit{ descriptive} model
of the phenomenon that Manski criticizes.

There have been recent attempts to model non-Bayesian researchers.
\citet{andrewsetal2021} show that conventional loss-minimizing estimators
may be suboptimal when consumers of the researcher are Bayesian with
heterogeneous priors. \citet{Banerjee2020TheoryExperimenters} describe
researchers as ambiguity averse max-minimizers. \citet{spiess2024}
models strategic choice of model misspecification by researchers.
In relation to this literature, our paper makes (to our knowledge)
the first attempt to model the role of assumptions in how researchers
interpret empirical observations.

\section{\label{sec:A-Model}A Model}

A \textit{representative researcher} is interested in a \textit{question}
whose answer is determined by \textit{fixed parameter}\textit{\emph{s}}
$\omega\in\Omega\subset\mathbb{R}^{n}$. The question is a subset
$Q\subseteq\{1,...,n\}$, indicating which of the parameters the researcher
wishes to learn. The researcher\textit{ }has a prior belief about
the fixed parameters. We assume that $\Omega$ is compact  and convex,
that her beliefs admit a continuous probability density $\mu$, and
that $\mu\left(\omega\right)=\prod_{i=1}^{n}\mu_{i}\left(\omega_{i}\right)>0$
for all $\omega\in\Omega$. We sometimes refer to $\mu$ as the researcher's
beliefs.

Time is discrete. In every period $t=1,2,...$, a real-valued vector
$\theta^{t}\in\Theta$ of \textit{context parameters}\textit{\emph{
is realize}}d. We refer to a realization of $\theta^{t}$ as a \textit{context}.
While $\omega$ represents a constant feature of the phenomenon of
interest (e.g., returns to education), $\theta^{t}$ represents transient,
circumstantial aspects of a particular dataset or experiment (e.g.,
whether assignment of students to educational treatments in a particular
setting is random). We assume that $\Theta$ is compact and convex,
and that $\theta^{t}$ admits a density $p_{\theta}$. 

In period $t$, the researcher observes $\theta^{t}$ and makes a
decision $a^{t}\in\left\{ 0,1\right\} $, indicating whether to conduct
research. If the researcher chooses $a^{t}=0$, then she passes over
the opportunity to conduct research. If research is conducted in period
$t$, then a vector of observed variables (referred to as \emph{statistics})
$s^{t}\in S$\textit{\emph{ and a vector of }}\emph{unobserved variables}\textit{\emph{
$u^{t}\in U$ }}are generated; both $S$ and $U$ are subsets of Euclidean
space. For expositional convenience, our definitions and general results
proceed as if $S$ and $U$ are both finite; extension to the continuum
case is straightforward and we make use of it in most of our examples.
Throughout, we adopt the notational convention that for a vector $x$,
$x_{B}=\left(x_{i}\right)_{i\in B}$ when $B$ is a subset of the
indices for $x$ and $x_{-i}=\left(x_{j}\right)_{j\neq i}$.

The data-generating process $p$ that governs the realization of $(u,s)$
at every time period satisfies $p\left(u^{t},s^{t}|\theta^{t},\omega\right)=p_{u}\left(u^{t}\right)p\left(s^{t}|u^{t},\theta^{t},\omega\right)$.
We assume that $p$ is continuous and has full support at every $\left(\theta,\omega\right)$.
Then, the density 
\[
p\left(s^{t},u^{t},\theta^{t},\omega\right)=\mu\left(\omega\right)p_{\theta}\left(\theta^{t}\right)p_{u}\left(u^{t}\right)p\left(s^{t}|u^{t},\theta^{t},\omega\right)
\]
describes her prior beliefs.\footnote{The density is the Radon-Nikodyn derivative with respect to the Lebesgue
measure on $\Theta\times\Omega$ multiplied by the uniform measure
on $S\times U$.} The context parameters and unobserved variables are distributed independently
and identically across periods.

An \textit{assumption} is an element $\theta^{*}\in\varTheta$. We
say that an assumption $\theta^{*}$ is \textit{identifying} for $Q$
if for every $\omega,\psi\in\Omega$ such that $\omega_{Q}\neq\psi_{Q}$,
there exists $s\in S$ such that $p\left(s\mid\theta^{*},\omega\right)\neq p\left(s\mid\theta^{*},\psi\right)$.
If an identifying assumption holds, then repeated observation of $s$
would eventually provide a definitive answer to the research question.
We assume that there is a single feasible identifying assumption.
In Section \ref{sec:Multiple/Structural}, we extend our model to
allow for multiple feasible identifying assumptions (including assumptions
about fixed parameters).

The researcher knows $p$. Entering period $t$, she observes the
current context $\theta^{t}$ and has beliefs about the fixed parameters
described by the density $\mu\left(\cdot|h^{t}\right)$, which depends
on the history $h^{t}=\left(a^{\tau},s^{\tau},\theta^{\tau}\right)_{\tau<t}$;
beliefs given the empty history equal the prior. There are a constant
$K$ and a divergence $D$ measuring how much one probability measure
on $S\times U$ deviates from another.\footnote{Recall that $D$ is a divergence if $D(m||m^{\prime})$ is positive,
is continuous in $\left(m,m^{\prime}\right)$, and equals $0$ if
and only if $m=m^{\prime}$ (a.s.).} Let $p_{S,U}\left(\cdot|\theta^{t},h^{t}\right)$ and $p_{S}\left(\cdot|\theta^{t},h^{t}\right)$
denote the researcher's beliefs given $\left(\theta^{t},h^{t}\right)$
over $\left(s^{t},u^{t}\right)$ and $s^{t}$, respectively. If
\[
D\left(p_{S,U}\left(\cdot|\theta^{t},h^{t}\right)||p_{S,U}\left(\cdot|\theta^{*},h^{t}\right)\right)>K,
\]
then the assumption is deemed implausible and the researcher chooses
$a^{t}=0$. Otherwise, the assumption is deemed sufficiently plausible
and the researcher chooses $a^{t}=1$. 

If the researcher chooses $a^{t}=0$, then she does not update her
beliefs, and so the next research opportunity, arising at period $t+1$,
is evaluated according to the same belief as in period $t$. If the
researcher chooses $a^{t}=1$, then she updates her belief over $\Omega$\textit{\emph{
}}\textit{as if}\textit{\emph{ the assumption $\theta^{*}$ }}held
exactly. That is, she now uses the density $\mu\left(\cdot|h^{t+1}\right)$
given by
\begin{equation}
\mu\left(\omega|h^{t},s^{t},a^{t}=1,\theta^{t}\right)=\frac{p\left(s^{t}|\theta^{*},\omega\right)\mu\left(\omega|h^{t}\right)}{p\left(s^{t}|\theta^{*},h^{t}\right)}\label{eq:belief updating}
\end{equation}
for almost every $\omega$. Denote by $\mu_{t}(\cdot)$ the Borel
probability measure with density $\mu\left(\cdot|h^{t}\right)$, noting
that $\mu_{t}$ is implicitly a function of $h^{t}$. Observe that
the dependence of $p_{S,U}\left(\cdot|\theta^{t},h^{t}\right)$ on
$h^{t}$ is restricted to its public part, namely $\left(s^{\tau}\right)_{\left\{ \tau:a^{\tau}=1\right\} }$,
because every time the researcher updates her belief, she does so
as if the context is $\theta^{*}$.

We assume that $D$ belongs to the class of $f$-divergences \citep{renyi1961entropy,csiszar1967information},
i.e., 
\[
D\left(m||m^{\prime}\right)=\sum_{s,u}m^{\prime}\left(s,u\right)f\left(\frac{m\left(s,u\right)}{m^{\prime}\left(s,u\right)}\right)
\]
for a strictly convex function $f\left(\cdot\right)$ satisfying $f(1)=0$.
In all of our examples, we take $D$ to be KL divergence, 
\[
D\left(m||m^{\prime}\right)=D_{KL}\left(m||m^{\prime}\right)=\sum_{s,u}m\left(s,u\right)\ln\left(\frac{m\left(s,u\right)}{m^{\prime}\left(s,u\right)}\right)
\]
the special case where $f\left(x\right)=x\ln x$.

\subsection{An example: A contaminated experiment\label{subsec:An-example:-A}}

To illustrate our model, consider a researcher who wants to identify
a behavioral effect from experimental data that is contaminated by
``friction.'' When the researcher conducts the experiment at period
$t$, she observes the statistic $s^{t}$ given by $s^{t}=\omega_{1}+\theta^{t}\omega_{2}+\varepsilon^{t}$.
She wants to learn the fixed parameter $\omega_{1}$, i.e., $Q=\{1\}$.
The fixed parameter $\omega_{2}$ represents the friction's strength.
The context $\theta\in[0,1]$ captures how well the experimental design
manages to curb the friction, and $\varepsilon^{t}\sim N(0,1)$ represents
noise and is independently drawn across periods. There are no latent
variables. The researcher's prior holds that each $\omega_{i}\sim N\left(m_{i},(\sigma_{i})^{2}\right)$,
independently of the other component of $\omega$. For example, $\omega_{1}$
is the degree of \textit{intrinsic }altruism in a certain social setting,
and $\omega_{2}$ is how much the subjects want an outside observer
to \textit{perceive} them as altruistic. The only identifying assumption
is $\theta^{*}=0$: under any $\theta^{t}\neq0$, both $\omega$ and
$(\omega_{1}-1,\omega_{2}+1/\theta^{t})$ generate the same distribution.

\subsection{Discussion}

The interpretation of the learning process is as follows. The researcher
can only update her beliefs under an identifying assumption, but will
do so only if she deems the assumption sufficiently plausible. We
refer to the decision to process the data at a given period as if
it is a decision whether to conduct the research at that period. This
fits an interpretation that the plausibility judgment is made by the
researcher herself. Alternatively, it could be viewed as a decision
by the \textit{research community} (embodied by seminar audiences
and journal referees) whether to ``take the research seriously''
and incorporate it into its collective knowledge. Under both interpretations,
the plausibility judgment at any given period is made \textit{before}
the research results are observed.

Plausibility is captured by how likely variable realizations are under
the actual context $\theta^{t}$ relative to the assumed one $\theta^{*}$.
The likelihood judgment is based on the researcher's current beliefs.
When relative likelihoods become farther away from $1$, the assumption
is deemed less plausible because it tends to produce unlikely predictions
about the variables, according to the researcher's belief. Convexity
of the function $f$ in an f-divergence means that the penalty for
a marginal shift of the likelihood ratio away from $1$ is increasing.

The plausibility judgment has additional noteworthy features. First,
it depends only on the current period's context and the current belief
$\mu_{t}$. Accordingly, the set of values of $\theta$ for which
the researcher conducts research depends on her belief $\mu_{t}$
and is denoted by $\text{\ensuremath{\varTheta^{R}\left(\mu_{t}\right)}}$.
Second, since $p\left(\cdot|\theta,\omega\right)$ has full support,
the data never definitively refute a wrong assumption. Consequently,
every assumption has finite f-divergence from the truth. Third, the
plausibility judgment takes into account the assumption's effect on
the distribution of \emph{both} observed ($s$) and latent ($u$)
variables. This reflects our observation of real-life discussions
of identification strategies in empirical economics. For instance,
evaluation of an instrumental variable is based on a judgment of whether
the (observed) instrument is correlated with (unobserved) confounding
variables. Finally, the constant $K$ captures the research community's
tolerance to implausible assumptions. While this tolerance can reflect
an underlying calculation of costs and benefit of doing research,
we do not explicitly model this calculus. Since the research community
knowingly chooses to distort its beliefs by making wrong assumptions,
it is not obvious how one should model such a cost-benefit analysis.

An alternative interpretation of the plausibility constraint is that
it reflects limitations on the ability to communicate the circumstances
of a scientific study across the research community. Under this interpretation,
a context parameter represents soft, high-dimensional information
that is difficult to communicate. The assumption represents an exogenous
communicability constraint. The researcher with the opportunity to
conduct research knows the true context (or has a good signal about
it) but cannot communicate it, yet she does not want to mislead the
community. The plausibility constraint captures the idea that she
only disseminates the research when she thinks it will not mislead
the community too much by ignoring the context.

Once the researcher in our model deems the assumption sufficiently
plausible, she updates her beliefs as if the assumption were perfectly
valid. Moreover, subsequent research never puts its plausibility in
doubt again. While admittedly extreme, we believe that this account
contains a kernel of truth, namely people's tendency to invoke ``working
hypotheses'' to facilitate information processing and later downplay
or forget the tentative nature of these hypotheses, effectively regarding
them as facts. It is also consistent with our casual observation that
debates over the adequacy of an identification strategy for a particular
study play an important role in the research community's decision
whether to take the study seriously (amplifying its exposure in seminars
and conferences, accepting it for publication in prestigious journals,
etc.), yet subsequent references to the published study rarely re-litigate
the identification strategy's appropriateness for that particular
study.

We conclude with a comment on the notion of identifying assumptions.
Recall that in the Contaminated Experiment example of Section \ref{subsec:An-example:-A},
we noted that the only feasible identifying assumption is $\theta^{*}=0$.
This claim rests on an implicit feature of our model: identification
assessments are made for each time period in isolation. Suppose we
observe the long-run distribution of $s$ for two known values of
$\theta$. Then, we have two equations with two unknowns ($\omega_{1}$
and $\omega_{2}$), and we can therefore pin down both. It follows
that if identification were assessed by combining multiple contexts
(given by different values of $\theta$), there would be no need for
identifying assumptions. This ``triangulating'' identification strategy
would work in most of the examples in this paper. However, it is natural
to assume that researchers cannot perform this kind of triangulation,
as they do not know future values of $\theta$ and past research opportunities
cannot be repeated. Moreover, triangulation is inconsistent with the
research practice we are familiar with, where the identification constraint
is applied separately to each piece of research.

\section{Research slowdown\label{sec:Speed}}

In this section we begin our analysis of the learning process spelled
out in the previous section. This section and the next are devoted
to the question of how the rate of conducting research evolves over
time.

\subsection{The Contaminated Experiment Continued\label{subsec:Contaminated-Experiments}}

Let us analyze learning dynamics for the example given by Section
\ref{subsec:An-example:-A}. Entering period $t$, the researcher
believes that $\omega_{i}\sim N(m_{i}^{t},(\sigma_{i}^{t})^{2})$
for each $i$. Whenever the researcher updates her beliefs, she does
so as if $\theta=0$, so her beliefs over $\omega_{2}$ never evolve;
accordingly, we remove the time index from the mean and variance of
$\omega_{2}$.

Entering period $t$, the researcher believes that the distribution
of $s^{t}$ conditional on $\theta^{t}$ is 
\[
N\left(m_{1}^{t}+\theta^{t}m_{2},1+\left(\sigma_{1}^{t}\right)^{2}+\left(\theta^{t}\right)^{2}\sigma_{2}^{2}\right).
\]
Using the standard formula for KL divergence between two scalar Gaussian
variables, 
\begin{align*}
D_{KL}\left(p_{S}\left(\cdot|h^{t},\theta^{t}\right)||p_{S}\left(\cdot|h^{t},\theta^{*}\right)\right) & =\frac{1}{2}\left[\left(\theta^{t}\right)^{2}\frac{\sigma_{2}^{2}+m_{2}^{2}}{1+\left(\sigma_{1}^{t}\right)^{2}}-\ln\left(1+\frac{\left(\theta^{t}\right)^{2}\sigma_{2}^{2}}{1+\left(\sigma_{1}^{t}\right)^{2}}\right)\right].
\end{align*}
Thus, the only time-varying elements that affect the propensity to
experiment are $\sigma_{1}^{t}$ and $\theta^{t}$.

The KL divergence is continuous and increasing in $\theta^{t}$, and
vanishes when $\theta^{t}=0$. Consequently, there exists a threshold
$\bar{\theta}\left(\sigma_{1}^{t}\right)>0$ such that the researcher
conducts research if and only if $\theta^{t}\in\left[0,\bar{\theta}\left(\sigma_{1}^{t}\right)\right]$.
Holding $\theta^{t}$ fixed, divergence decreases in $\sigma_{1}^{t}$,
so the threshold for conducting research $\bar{\theta}\left(\cdot\right)$
increases in $\sigma_{1}^{t}$.

When the researcher conducts the experiment in period $t$, the variance
of her belief about $\omega_{1}$ decreases to $\sigma_{1}^{t+1}=\sigma_{1}^{t}\left(\left(\sigma_{1}^{t}\right)^{2}+1\right)^{-\frac{1}{2}}$.
That is, $\sigma_{1}^{t}$ decreases monotonically over time. Consequently,
the propensity to conduct research uniformly decreases over time.
As the researcher becomes more certain of her belief over $\omega_{1}$,
she also becomes more sensitive to the noise and so more reluctant
to assume it away. In other words, her standards for what passes as
adequate research design increase over time. This slows down the rate
of learning.

However, learning takes place with positive frequency in the long
run. To see why, note that as $\sigma_{1}^{t}\rightarrow0$, the divergence
converges to
\[
\frac{1}{2}\left[\left(\sigma_{2}^{2}+m_{2}^{2}\right)\left(\theta^{t}\right)^{2}-\ln\left(1+\left(\theta^{t}\right)^{2}\sigma_{2}^{2}\right)\right],
\]
which is finite for every $\theta^{t}$. This means that $\bar{\theta}\left(0\right)>0$,
and research takes place with positive probability, regardless of
the researcher's current belief. This non-vanishing learning implies
that $\sigma_{1}^{t}\rightarrow0$ as $t\rightarrow\infty$. In the
long-run, research is carried out whenever $\theta\in\left[0,\bar{\theta}\left(0\right)\right]$,
and that the researcher's belief over $\omega_{1}$ assigns probability
one to
\[
\omega_{1}+\mathbb{E}\left(\theta|\theta<\bar{\theta}\left(0\right)\right)\omega_{2}.
\]
Thus, the long-run estimate of the effect of interest is biased in
proportion to the true value of the friction parameter $\omega_{2}$.
The magnitude of the bias also increases with $\sigma_{2}^{2}$ (the
researcher's time-invariant uncertainty over the friction parameter)
since $\bar{\theta}\left(0\right)$ increases with $\sigma_{2}^{2}$.

To summarize our findings in this example, the researcher's propensity
to learn decreases over time but remains positive in the long-run.
This in turn means that the long-run answer to the research question
is biased. The bias is proportional to the true value of the fixed
friction parameter, and increases (in absolute terms) with the researcher's
uncertainty over it.

\subsection{General results}

In the above example, the set of contexts for which research takes
place (weakly) contracts over time. This shows the possibility of
a uniform (i.e., with probability one) research slowdown. Our first
two results show that the opposite pattern, namely a uniform increase
in the propensity to conduct research, cannot occur. Consequently,
the rate of research decreases at least with some probability.
\begin{prop}
\label{prop: non-expansion}For any $\theta\in\Theta$ and history
$h^{t}$, if $\mathbb{P}\left[\theta\in\Theta^{R}\left(\mu_{t+1}\right)\setminus\Theta^{R}\left(\mu_{t}\right)|h^{t}\right]>0$,
then there exists $t^{*}>t+1$ such that $\mathbb{P}\left[\theta\notin\varTheta^{R}\left(\mu_{t^{*}}\right)|h^{t+1}\right]>0$
.
\end{prop}
This result states that any expansion in the set of parameters for
which research is conducted reverses itself with positive probability.
Consider a context for which research does not takes place in some
period. Suppose that there is some piece of evidence that would lead
to research being performed for that same context in the following
period. The result shows that with positive probability, there is
a point in the future at which the researcher would once again refrain
from conducting research in that same context.

When a larger difference in context maps naturally to a bigger divergence,
we can be more explicit about how the propensity to conduct research
evolves.
\begin{prop}
\label{prop: quasi-convex}Suppose that $D\left(p_{S,U}\left(\cdot|\theta,h^{t}\right)||p_{S,U}\left(\cdot|\theta^{*},h^{t}\right)\right)$
is quasi-convex in $\theta$ for every history $h^{t}$. If 
\[
\mathbb{P}\left[\Theta^{R}\left(\mu_{t+1}\right)\setminus\Theta^{R}\left(\mu_{t}\right)\neq\emptyset|h^{t}\right]>0,
\]
then
\[
\mathbb{P}\left[\Theta^{R}\left(\mu_{t}\right)\setminus\Theta^{R}\left(\mu_{t+1}\right)\neq\emptyset|h^{t}\right]>0.
\]
\end{prop}
This result says that when there are contexts for which research takes
place at period $t+1$ but not at $t$ (i.e., $\Theta^{R}\left(\mu_{t+1}\right)\setminus\Theta^{R}\left(\mu_{t}\right)\neq\emptyset$),
then with positive probability, there are contexts for which research
takes place at $t$ but not at $t+1$ (i.e., $\Theta^{R}\left(\mu_{t}\right)\setminus\Theta^{R}\left(\mu_{t+1}\right)\neq\emptyset$).
That is, at any history, when the community conducts research in new
contexts with positive probability, it also stops conducting research
in others. When one value of the statistic increases the chance of
conducting research, a different value decreases the chance.

The result relies on the assumption that the divergence is quasi-convex
in $\theta$. In particular, this holds when a larger Euclidean distance
between $\theta$ and $\theta^{*}$ implies a larger divergence. In
our examples, $\theta\in\mathbb{R}_{+}$, $\theta^{*}=0$, and the
divergence strictly increases in $\theta$. Consequently, Proposition
\ref{prop: quasi-convex} applies to all of our examples.

The proofs of Propositions \ref{prop: non-expansion} and \ref{prop: quasi-convex}
both rely on convexity of f-divergence. This implies that the divergence
of $p_{S,U}(\cdot|\theta,h^{t})$ from $p_{S,U}(\cdot|\theta^{*},h^{t})$
goes up in expectation for every $\theta$ when $h^{t}$ is concatenated
by additional observations. If it decreases for some histories, then
it must rise for others. Both proofs exploit this insight to show
that expansions in $\Theta^{R}$ must be offset by contractions in
it.

\section{Constant Propensity to Research\label{subsec:Constant research}}

In this section we continue to explore the rate of research in our
model, focusing on cases in which this rate remains constant over
time.

\subsection{An Example: Confounded causal inference}

Determining the causal effect of one variable on another is a central
task for empirical researchers. To do so, researchers must account
for (unobserved) confounding variables that affect both the (observed)
cause and effect. We present a stylized example of causal inference
from observational data in the presence of a potential confounder.

There are two observable variables, a potential cause $s_{1}$ and
an outcome $s_{2}$. The researcher wants to learn the causal effect
of the former on the latter. This effect is parameterized by $\omega_{1}\in(-1,1)$,
i.e., $Q=\{1\}$. However, the observed correlation between the two
variables is confounded by a latent variable $u$ that affects both.
The fixed parameter $\omega_{2}\in(-1,1)$ captures the strength of
this confounding effect. The context parameter $\theta\in[0,1]$ captures
the extent to which it persists in a given dataset. A lower value
of $\theta$ captures better ``research design'' in the sense of
attenuating the confounding effect. If she conducts research in period
$t$, then she observes 
\begin{align*}
s_{1}^{t} & =\theta^{t}\omega_{2}u^{t}+\varepsilon_{1}^{t}\\
s_{2}^{t} & =\omega_{1}s_{1}^{t}+\omega_{3}u^{t}+\varepsilon_{2}^{t}
\end{align*}
where $u^{t}\sim N\left(0,1\right)$ and $\varepsilon_{i}^{t}\sim N\left(0,\sigma_{i,t}^{2}\right)$
for $i=1,2$, independently of each other. There is no uncertainty
regarding $\omega_{3}>0$. Set this parameter and the variances $\sigma_{1,t}^{2}$
and $\sigma_{2,t}^{2}$ such that $s_{i}|\omega,\theta^{t}\sim N\left(0,1\right)$
for each $i=1,2$ and $\left(\theta,\omega\right)$.\footnote{That is, $\sigma_{1,t}^{2}=1-(\theta^{t})^{2}\omega_{1}^{2}$ and
$\sigma_{2,t}^{2}=1-\omega_{2}^{2}-\omega_{3}^{2}-2\theta^{t}\omega_{1}\omega_{2}\omega_{3}$.} It follows that the only aspect of the long-run distribution of $s^{t}=\left(s_{1}^{t},s_{2}^{t}\right)$
that sheds light on the fixed parameters is the pairwise correlation
between the two statistics,
\[
\rho_{12}\left(\theta^{t},\omega\right)=(\theta^{t})^{2}\omega_{2}^{2}\omega_{2}+\theta^{t}\omega_{2}\omega_{3}+\omega_{1}.
\]
The equation for $\rho_{12}\left(\theta^{t},\omega\right)$ reveals
that the only identifying assumption is $\theta^{*}=0$. Under that
assumption, the correlation between $s_{1}$ and $s_{2}$ equals $\rho_{12}\left(\theta^{*},\omega\right)=\omega_{1}$
and so pins down the causal effect of interest. Note that, as in the
previous example, this assumption prevents the researcher from learning
anything about the other fixed parameter ($\omega_{2}$).

We derive an expression for the KL divergence between the true and
assumed distributions over $(u,s)$. The joint density of the statistics
conditional on the parameters factorizes as $p(u,s|\omega,\theta)=p(u)p(s_{1}|u,\omega_{1},\theta)p(s_{2}|u,s_{1},\omega_{2})$.
It follows that

\begin{align*}
 & D_{KL}\left(p_{S,U}\left(\cdot|\theta^{t},h^{t}\right)||p_{S,U}\left(\cdot|\theta^{*},h^{t}\right)\right)\\
= & \int\ln\left(\frac{\int p(u)p(s_{1}|u,\omega_{2},\theta^{t})p(s_{2}|u,s_{1},\omega_{1})\mu\left(\omega_{1},\omega_{2}|h^{t}\right)d\omega}{\int p(u)p(s_{1}|u,\omega_{2},\theta^{*})p(s_{2}|u,s_{1},\omega_{1})\mu\left(\omega_{1},\omega_{2}|h^{t}\right)d\omega}\right)dp\left(s,u|\theta^{t}\right)\\
= & \int\ln\left(\frac{\int p\left(s_{1}|u,\omega_{2},\theta^{t}\right)\mu\left(\omega_{2}|h^{t}\right)d\omega_{2}\int p\left(s_{2}|s_{1},u,\omega_{1}\right)\mu\left(\omega_{1}|h^{t}\right)d\omega_{1}}{\int p\left(s_{1}|u,\omega_{2},\theta^{*}\right)\mu\left(\omega_{2}|h^{t}\right)d\omega_{2}\int p\left(s_{2}|s_{1},u,\omega_{1}\right)\mu\left(\omega_{1}|h^{t}\right)d\omega_{1}}\right)dp\left(s,u|\theta^{t}\right)\\
= & \int\ln\left(\frac{\int p\left(s_{1}|u,\omega_{2},\theta^{t}\right)\mu\left(\omega_{2}\right)d\omega_{2}}{\int p\left(s_{1}|u,\omega_{2},\theta^{*}\right)\mu\left(\omega_{2}\right)d\omega_{2}}\right)dp\left(s_{1},u|\theta^{t}\right)
\end{align*}
Note that only the researcher's belief about $\omega_{2}$ enters
the expression. Because this belief is stationary, the divergence
for any given $\theta^{t}$ does not change over time. Therefore,
the researcher's propensity to research is time-invariant: there is
a fixed $\bar{\theta}$ such that she conducts research if and only
if $\theta^{t}\in\left[0,\bar{\theta}\right]$. As $t\rightarrow\infty$,
the researcher's belief is concentrated on
\[
\widehat{\omega}_{2}=\mathbb{E}\left[\rho_{12}(\theta,\omega)|\theta<\bar{\theta}\right].
\]
Clearly, this long-run estimate is biased when $\omega_{2}\neq0$,
i.e., when there is a confounding effect.

\subsection{A General Result}

In the above example, the set of contexts for which research takes
place is history-independent. Our next result provides a general sufficient
condition for this property. The condition concerns the probabilistic
relationship between the context and the fixed parameters about which
the researcher learns under the identifying assumption. We call these
parameters \textit{active}\textit{\emph{, and formally define the
set of}}\textit{ active parameters} $Q^{*}$ to be the smallest set
of indices for which $p_{S}\left(\cdot|\theta^{*},\omega\right)=p_{S}\left(\cdot|\theta^{*},\omega^{\prime}\right)$
whenever $\omega_{Q^{*}}=\omega_{Q^{*}}^{\prime}$. Under the identifying
assumption, no other fixed parameters affect the long-run distribution
of $s$, and so repeated observation teaches the researcher nothing
about them. The set of active parameters is defined with respect to
$\theta^{*}$, and by the definition of identifying assumptions, $Q\subseteq Q^{*}$.
In both examples, the set of active parameters was $Q^{*}=Q=\{1\}$,
though this need not be the case (see Section \ref{subsec: Instrumental-variabl}).

Our result makes use of the underlying recursive structure of the
data-generating process as described by a directed acyclic graph (DAG),
a graph whose edges correspond to an acyclic binary relation. Our
analysis employs tools from the AI/Statistics literature on DAGs (see
\citet{pearl2009} or \citet{kollerFriedman2009} for a general introduction,
and \citet{spiegler2016,Spiegler2020Review} or \citet{ellisthysen2022}
for earlier economic-theory applications). Denote $x^{t}=\left(s^{t},u^{t},\theta^{t},\omega\right)\in\mathbb{R}^{m}$,
$N=\left\{ 1,\dots,m\right\} $, and $N^{s}\subset N$ the set of
indices such that $x_{N^{s}}^{t}=s^{t}$.\footnote{The components of $x^{t}$ are ordered such that $s_{i}^{t}=x_{i}^{t}$.}
Let $G=(N,R)$ be a DAG with nodes $N$ and set of directed edges
$R$ such that no edge goes into any node in $N\setminus N^{s}$.
We say that $p$ has the\emph{ recursive structure} $G$ if the density
of $x^{t}=\left(s^{t},u^{t},\theta^{t},\omega\right)$ can be written
as
\begin{equation}
\mu\left(\omega\right)p_{\theta}\left(\theta^{t}\right)p_{u}\left(u^{t}\right)\prod_{i\in N^{s}}p\left(s_{i}^{t}|x_{R(i)}^{t}\right)\label{eq:DAG}
\end{equation}
for every $x^{t}$. In this formula, $R(i)$ is the set of nodes that
send directed edges into $i$, with customary abuse of notation.

Every $p$ has a recursive structure (in particular, one where every
statistic node is linked to every other node, in which case Equation
(\ref{eq:DAG}) holds by a standard chain rule). However, sparser
structures describe essential features of the data-generating process.
Whenever $p$ is described by a recursive system of regression equations,
as in all the examples in this paper, this system defines a recursive
structure for $p$: for every $i\in N^{s}$, $R(i)$ is the set of
R.H.S. variables in the equation for $s_{i}$. For instance, the recursive
structure of $p$ in the Causal Inference example is
\[
\begin{array}{ccccc}
\theta^{t} & \rightarrow & s_{1}^{t} & \leftarrow & \omega_{2}\\
 & \nearrow & \downarrow\\
u^{t} & \rightarrow & s_{2}^{t} & \leftarrow & \omega_{1}
\end{array}.
\]

A DAG $G$ satisfies a conditional-independence property if every
probability measure that is consistent with $G$ satisfies this property.
We say that $\theta$ and $\omega_{Q^{*}}$ are $G$\emph{-separable
}if for every $i$, $G$ satisfies $s_{i}^{t}\perp\omega_{Q^{*}}$
whenever it satisfies $s_{i}^{t}\not\perp\theta^{t}|\left(s_{-i}^{t},u^{t}\right)$.
If $\theta$ and $\omega_{Q^{*}}$ are $G$-separable, then the active
parameters do not influence any statistic whose distribution depends
on the context (conditional on the other variables). Thus, G-separability
describes a sense in which context and active parameters have distinct
observable effects. Structural conditional-independence properties
such as $G$-separability have a graphical characterization known
as ``d-separation'' (see \citet{pearl2009}), which makes the condition
easy to check. The Contaminated Experiment example violates $G$-separability,
since the context parameter and the fixed parameter of interest both
send links to the statistic. In contrast, the Causal Inference example
satisfies the property. The context parameter $\theta^{t}$ is d-separated
from $s_{2}^{t}$ given $(s_{1}^{t},u^{t})$ since the latter block
every path from $\theta^{t}$ to $s_{2}^{t}$, hence $s_{i}^{t}\perp\omega_{Q^{*}}|(s_{1}^{t},u^{t})$.
The fixed parameter $\omega_{1}=\omega_{Q^{*}}$ and $s_{1}^{t}$
are independent because they have no common ancestor.
\begin{prop}
\label{prop: constant rage-1} If $p$ has a recursive structure $G$
for which $\theta$ and $\omega_{Q^{*}}$ are $G$-separable, then
\textup{$\Theta^{R}\left(\cdot\right)$} is constant.
\end{prop}
The conditional-independence property that underlies Proposition \ref{prop: constant rage-1}
is not imposed directly on the researcher's belief, but rather on
its underlying structure $G$. Under \textit{G}-separability, the
set of contexts in which the researcher updates does not change over
time. That is, when context and active parameters have separate observable
effects, the propensity to conduct research is constant. The proof
uses d-separation to factorize beliefs into conditional-probability
terms. We show that every statistic whose distribution is sensitive
to the identifying assumption must be conditionally independent of
the active parameters. This in turn implies that every term involving
$\omega_{Q^{*}}$ cancels out or integrates out of the expression
for the f-divergence. Since the researcher's beliefs about the other
fixed parameters do not change, the divergence for any given context
remains fixed over time.

Proposition \ref{prop: constant rage-1} relies on $D$ being an expectation
of some function of likelihood ratios (induced by the true and assumed
values of $\theta$). However, it does not rely on the convexity of
$f$. In contrast, Propositions \ref{prop: non-expansion} and \ref{prop: quasi-convex}
rely on convexity of $D$, which follows from the convexity of $f$.

An identifying assumption makes $G$-separability non-trivial. If
all fixed parameters are active (as would be the case for most values
of $\theta$), then $G$-separability requires that when a statistic
depends on a context parameter, it must be independent of all fixed
parameters, which is an extreme requirement that trivializes the problem.
However, identifying assumptions typically imply the existence of
inactive fixed parameters. Identification problems arise when the
number of fixed parameters exceeds the number of ``moments'' that
the researcher can extract from the statistic. This leads to a system
of equations with too many unknowns. An identifying assumption effectively
shuts down some of these unknowns. In our language, these shutdown
unknowns are inactive parameters. When there are such inactive parameters,
$G$-separability is not a trivially demanding condition.

\subsection{\label{subsec: Instrumental-variabl}An Application: Instrumental
Variables}

Proposition \ref{prop: constant rage-1} allows a convenient analysis
of whether the propensity to adopt identification strategies for causal
inference changes over time. Consider a data-generating process described
by the following recursive system of regression equations:
\begin{eqnarray*}
s_{1}^{t} & = & \omega_{1}\theta^{t}u^{t}+\varepsilon_{1}^{t}\\
s_{2}^{t} & = & \omega_{2}s_{1}^{t}+\omega_{3}u^{t}+\varepsilon_{2}^{t}\\
s_{3}^{t} & = & \omega_{4}s_{2}^{t}+\omega_{5}u^{t}+\varepsilon_{3}^{t}
\end{eqnarray*}
where $u^{t}$ and the $\varepsilon_{i}^{t}$ variables are all independent
Gaussians. Their variances and the range of possible values of the
parameters are set so that $s_{i}^{t}|\left(\theta^{t},\omega\right)\sim N(0,1)$
for every $i=1,2,3$.

The statistic $s_{2}^{t}$ represents a potential cause of $s_{3}^{t}$,
which represents an outcome. The statistic $s_{1}^{t}$ is a potential
instrument. The researcher wants to learn $\omega_{4}$, the causal
effect of $s_{2}^{t}$ on $s_{3}^{t}$, i.e. $Q=\{4\}$. The latent
variable $u^{t}$ obfuscates the causal effect because it influences
$s_{1}^{t}$, $s_{2}^{t}$, and $s_{3}^{t}$. Since the statistic
variables are all standard normal, the only aspects of the long-run
distribution of $s^{t}$ that the researcher can use to learn $\omega$
are $E\left[s_{1}^{t}s_{2}^{t}\right]$, $E\left[s_{2}^{t}s_{3}^{t}\right]$
and $E\left[s_{1}^{t}s_{3}^{t}\right]$. This gives three equations
with five unknowns, and so $\omega_{4}$ cannot be identified. However,
when we make the assumption $\theta^{*}=0$, we get $\omega_{4}=E\left[s_{1}^{t}s_{3}^{t}\right]/E\left[s_{1}^{t}s_{2}^{t}\right]$.
This is the textbook 2SLS procedure that uses $s_{1}^{t}$ as an instrument
for $s_{2}^{t}$. The identifying assumption is that the instrument
is independent of the confounding variable $u^{t}$.

We can apply Proposition \ref{prop: constant rage-1} to this example.
The process has a structure given by the DAG
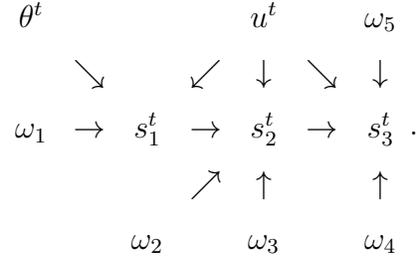
\[
\begin{array}{ccccccc}
\theta^{t} &  &  &  & u^{t} &  & \omega_{5}\\
 & \searrow &  & \swarrow & \downarrow & \searrow & \downarrow\\
\omega_{1} & \rightarrow & s_{1}^{t} & \rightarrow & s_{2}^{t} & \rightarrow & s_{3}^{t}\\
 &  &  & \nearrow & \uparrow &  & \uparrow\\
 &  & \omega_{2} &  & \omega_{3} &  & \omega_{4}
\end{array}.
\]
The active parameters are $\omega_{2}$, $\omega_{3}$, $\omega_{4}$
and $\omega_{5}$, i.e., $Q^{*}=\left\{ 2,3,4,5\right\} $, and $\theta$
and $\omega_{Q^{*}}$ are $G$-separable. Only $s_{1}^{t}$ is not
independent of $\theta^{t}$ conditional on the other variables: the
DAG includes a link $\theta^{t}\rightarrow s_{1}^{t}$, and using
d-separation, one can show that the other two statistics, $s_{2}^{t}$
and $s_{3}^{t}$, are both independent of $\theta^{t}$ given $s_{1}^{t}$
and $u^{t}$. However, $s_{1}^{t}$ is independent of $\omega_{Q^{*}}$
since they have no common ancestor. By Proposition \ref{prop: constant rage-1},
the researcher's propensity to employ the instrumental-variable identification
strategy is time-invariant.\footnote{In a previous version of the paper (\citet{EllisSpiegler2024}), we
examined another causal-inference identification strategy, known as
``front door identification'' (see \citet{pearl2009}), and showed
that it violates the condition for time-invariant propensity to learn.}

\section{Limiting Beliefs\label{sec:Limiting Beliefs}}

We now turn to the question of what the research community believes
about the fixed parameters of interest in the long run. We begin with
a definition of stable beliefs. Convergence of beliefs is according
to the weak{*} topology.
\begin{defn}
A Borel measure $\mu^{*}$ is \emph{stable} for $\omega^{*}$ if $\mathbb{P}\left(\mu_{t}\rightarrow\mu^{*}|\omega^{*}\right)>0$.
\end{defn}
A belief is stable when the posterior beliefs generated by the learning
process converge to it with positive probability in the long run.
Similar definitions appear in, e.g., \citet{FudenbergKreps1993} and
\citet{espondaPouzo2016}. In all our prior examples, unique stable
beliefs exist for all values of the fixed parameters, and the research
process converges to that stable belief with probability one.

\subsection{A General Characterization}

The following result characterizes stable beliefs.
\begin{prop}
\label{prop: long run beliefs}For any $\omega^{*}\in\Omega$, if
$\mu^{*}$ is stable for $\omega^{*}$ and $\Theta^{R}\left(\cdot\right)$
is continuous in a neighborhood of $\mu^{*}$, then $\mu^{*}\left(O\right)=1$
for any open $O\subset\Omega$ with
\begin{equation}
O\supset\arg\min_{\omega^{\prime}\in\Omega}D_{KL}\left(p_{S}\left(\cdot|\theta\in\varTheta^{R}\left(\mu^{*}\right),\omega^{*}\right)||p_{S}\left(\cdot|\theta^{*},\omega^{\prime}\right)\right).\label{eq:KL for stable}
\end{equation}
\end{prop}
Recall that observed statistics are affected both by the contexts
in which research is conducted and the true value of the fixed parameters,
$\omega^{*}$. If the researcher consistently holds a belief close
to $\mu^{*}$ for a long stretch of time, then the set of values of
$\theta$ for which research takes place during that stretch is close
to $\varTheta^{R}\left(\mu^{*}\right)$ and the long-run frequency
of the statistic is close to $p_{S}\left(s|\theta\in\varTheta^{R}\left(\mu^{*}\right),\omega^{*}\right)$.
However, the researcher updates her belief as if the context parameter
is $\theta^{*}$, and under that assumption, $s$ occurs with probability
$p_{S}\left(s|\theta^{*},\omega\right)$ with parameter $\omega$.
Following \citet{Berk1966} and \citet{espondaPouzo2016}, the long-run
belief that emerges from this misspecified Bayesian learning rules
out parameter values that do not minimize the KL divergence from the
true distribution to the subjective one. Therefore stable beliefs
attach positive probability only to the parameters close to those
that minimize this divergence.

We should not confuse the KL divergence in the result with the divergence
governing the decision whether to conduct research. In the latter
case, the divergence plays a similar role to a utility function that
captures the researcher's preferences and dictates her actions at
each period. In the former case, it is a statistical property of the
long-run empirical frequency of the observable variables. 

In what follows, we assume that every primitive function (e.g. $\mu\left(\cdot\right)$
and $f\left(\cdot\right)$) is smooth, that $\Theta=[0,1]$, and that
$\theta^{*}=0$. Our next result studies whether the stable beliefs
are correct. We show they are not for ``most'' models in the following
class, where a data-generating process $p$ and cutoff $K$ constitute
a model. 

We say that a model $\left(p,K\right)$ is \emph{regular} if \emph{(i)}
$Q^{*}\neq\left\{ 1,\dots,n\right\} $; \emph{(ii) }the divergence
\[
D_{KL}\left(\int p_{S,U}\left(\cdot|\theta,\omega\right)dm||\int p_{S,U}\left(\cdot|\theta^{*},\omega\right)dm\right)
\]
is strictly increasing in $\theta$ for every probability measure
$m$ on $\Omega$; and \emph{(iii)} for every $\omega$ and some $j\in Q$,
$\frac{d}{d\omega_{j}}p_{S}(s|\omega,\theta^{*})\neq0$ for some $s$.
All of our examples have regular models. The first requirement is
that the assumption shuts down at least one fixed parameter. Therefore,
the assumption disregards some parameters that affect the statistics.
The second requires that the context parameter captures how far the
assumed distribution is from the realized distribution when distance
is captured by $D\left(\cdot\right)$. This implies that there exist
a function $\bar{\theta}\left(\cdot\right)>0$ so that $\Theta^{R}\left(\mu_{t}\right)=\left[0,\bar{\theta}\left(\mu_{t}\right)\right]$.
The function $\bar{\theta}\left(\mu_{t}\right)$ is continuous, so
we can apply Proposition \ref{prop: long run beliefs}. The third
requires that the distribution of statistics is sensitive to some
parameter $j\in Q$ under $\theta^{*}$.

In all of our examples, the answer to the question at which the researcher
eventually arrives is incorrect with probability one. For every $\omega^{*}$
with $\omega_{2}^{*}\neq0$, the researcher almost surely comes to
believe that $\omega_{Q}$ equals $\hat{\omega}_{Q}\neq\omega_{Q}^{*}$
with certainty. That is, researcher's use of an identifying assumption
leads to wrong beliefs and what Manski terms ``incredible certitude.''
We say that a model exhibits \emph{incredible certitude} if for almost
every $\omega^{*}\in\Omega$, every stable belief $\mu^{*}$ for $\omega^{*}$
attaches probability one to some $\hat{\omega}_{Q}\neq\omega_{Q}^{*}$.
\begin{prop}
\label{prop: generic incredible certitude}Suppose that $D_{KL}\left(p_{S}\left(\cdot|\theta,\omega^{*}\right)||p_{S}\left(\cdot|\theta^{*},\omega^{\prime}\right)\right)$
is strictly convex in $\omega_{Q^{*}}^{\prime}$ for any $\theta\in\Theta$
and any $\omega^{*}\in\Omega$. Then, every stable belief $\mu^{*}$
for any $\omega^{*}\in\Omega$ has a degenerate marginal on $\omega_{Q}$
when $\left(p,K\right)$ is regular. Moreover, the models that exhibit
incredible certitude are dense within the set of regular models.
\end{prop}
The result shows that two features of the examples lead to incredible
certitude. First, the KL divergence between the empirical distribution
of statistics in $\omega^{*}$ and the distribution of statistics
in state $\omega^{\prime}$ under the identifying assumption is strictly
convex in $\omega_{Q^{*}}^{\prime}$. Since KL divergence is a convex
function of probability measures, this holds when the assumed distribution
resulting from $\alpha\omega^{\prime\prime}+(1-\alpha)\omega^{\prime}$
is in between the ones resulting from $\omega^{\prime}$ and $\omega^{\prime\prime}$.
This ensures that every stable belief is degenerate. Second, the examples
have regular models. 

The proof of the proposition shows that for any $Q^{*}$, almost all
small perturbations of a regular model (within a class that leaves
$Q^{*}$ and $\theta^{*}$ unchanged) exhibit incredible certitude.\footnote{Specifically, the perturbations are changes in $K$ or adding any
polynomial (with small coefficients) to the data-generating process
that does not change $Q^{*}$.} We construct a necessary condition for a fixed parameter to lead
to correct beliefs. Then, we apply the transversality theorem to a
parameterization of the set of perturbations. This establishes that
almost all of them do not satisfy it for a full measure of states.

\subsection{\label{subsec:Multiple}An Example: Contaminated experiments, revisited}

There may be multiple stable beliefs. In this subsection, we demonstrate
this possibility using a variant on the example from Section \ref{subsec:Contaminated-Experiments}.
Let $s^{t}\in\left\{ 0,1\right\} $, $\omega=\left(\omega_{1},\omega_{2}\right)\in\Omega=\left[\varepsilon,1-\varepsilon\right]^{2}$
for $\varepsilon\in\left(0,\frac{1}{2}\right)$, and $\theta\in[0,1]$.
The data-generating process is
\[
p\left(s^{t}=1\mid\theta^{t},\omega\right)=\left(1-\theta^{t}\right)\omega_{1}+\theta^{t}\omega_{2},
\]
and the researcher wants to learn $\omega_{1}$, i.e., $Q=\{1\}$.
The only identifying assumption is $\theta^{*}=0$, which prevents
learning anything about $\omega_{2}$. The divergence for the assumption
is an increasing function of $\bar{\mu}_{1}^{t}\equiv\mathbb{\mathbb{E}}\left[\omega_{1}|h^{t}\right]$,
and the researcher imposes the assumption whenever $\theta^{t}\leq\bar{\theta}\left(\bar{\mu}_{1}^{t}\right)$
for some threshold $\bar{\theta}\left(\cdot\right)>0$.

Suppose that $\mathbb{E}_{\mu}\left[\omega_{2}\right]=\frac{1}{2}$,
and that the distribution over $\theta$ is smooth with full support
and mean $\frac{1}{2}$. Any candidate for a stable belief given the
true $\omega^{*}$ assigns probability one to a $\omega_{1}$ that
satisfies
\begin{equation}
\omega_{1}=E\left[\theta|\theta\leq\bar{\theta}(\omega_{1})\right]\cdot\left(\omega_{2}^{*}-\omega_{1}^{*}\right)+\omega_{1}^{*}\label{eq: steady-state equation}
\end{equation}
by Proposition \ref{prop: long run beliefs}. The next result shows
that Equation (\ref{eq: steady-state equation}) has at least two
solutions when $K$ is small.
\begin{prop}
\label{prop: multiple steady states}For any $K>0$, $a_{1}\in\left(\frac{1}{2},1-\varepsilon\right)$,
and $\zeta>0$, there exists $\delta>0$ so that for every $\omega^{*}\in B_{\delta}\left(\left(a_{1},1-a_{1}\right)\right)$,
there exists an attracting solution to Equation (\ref{eq: steady-state equation})
in $\left(\frac{1-\zeta}{2},\frac{1+\zeta}{2}\right)$. Moreover,
if $K$ and $\zeta$ are sufficiently small, this $\delta$ can be
chosen so that there also exists an attracting solution to Equation
(\ref{eq: steady-state equation}) that is greater than $\frac{1+\zeta}{2}$.
\end{prop}
The result establishes the possibility of multiple stable beliefs.
Moreover, even if the threshold $K$ is small, there may be stable
beliefs far from the truth. For any realized $\omega^{*}$, including
one with $\omega_{1}^{*}$ close to $1-\varepsilon$, there is a stable
belief that attaches probability one to a $\omega_{1}$ close to $\frac{1}{2}$,
regardless of the size of $K$. Put another way, the researcher becomes
certain that the question's answer is about $\frac{1}{2}$ when it
instead is close to $1-\varepsilon$, the maximum.

Notice that when research is conducted in period $t$, $\bar{\mu}_{1}^{t+1}>\bar{\mu}_{1}^{t}$
whenever $s^{t}=1$ and $\bar{\mu}_{1}^{t+1}<\bar{\mu}_{1}^{t}$ whenever
$s^{t}=0$. Since the threshold $\bar{\theta}$ is a strictly increasing
function of $\bar{\mu}_{1}^{t}$, there will be phases of both accelerating
and decelerating rates of research. This is in contrast to the uniform
research slowdown that emerged when the statistic was Gaussian.

\section{Assumption-Based Learning vs. (Misspecified) Bayesian Learning}

This section discusses how assumption-based learning differs from
Bayesian learning, both with and without misspecified prior beliefs.
Bayesian learning corresponds to always taking $a^{t}=1$ and updating
according to Bayes rule, i.e. $\mu\left(\omega|h^{t},s^{t}\right)p\left(s^{t}|h^{t},\theta^{t}\right)=\mu\left(\omega|h^{t}\right)p\left(s^{t}|\omega,\theta^{t}\right)$.
Misspecified Bayesian learning is the limiting case of our model where
$K=\infty$ and $\mu\left(\omega\right)p_{u}(u^{t})p(s^{t}|\theta^{*},u^{t},\omega)$
is the misspecified prior. We draw a contrast between the two and
assumption-based learning. 

In the most comparable interpretation of misspecified Bayesian learning,
a researcher makes an assumption (before the learning begins) that
determines her prior beliefs about the data-generating process. She
then imposes it at every research opportunity without reconsidering
its appropriateness. Under assumption-base learning, the researcher
re-evaluates the assumption's plausibility in every period against
her current beliefs and the current context. The latter generates
non-trivial predictions regarding the evolution of the rate of learning.
Moreover, stable beliefs are an equilibrium object, as evident from
the characterization in Proposition \ref{prop: long run beliefs}.
Under misspecified Bayesian learning, experiments take place in every
period, so all equilibrium effects vanish. Stable beliefs are still
given by (\ref{eq:KL for stable}) with $\Theta^{R}(\mu^{*})=\Theta$
for every $\mu^{*}$, which delivers the original characterization
by \citet{Berk1966}.

This highlights a difference in the assumptions' epistemic status
between assumption-based and misspecified Bayesian learning. In the
former, assumptions do not reflect the researcher's genuine beliefs.
Instead, they are ``working hypotheses'' in service of her mission
to infer the fixed parameters from a particular dataset. For instance,
in the Contaminated Experiment example, the researcher does not assume
$\theta^{*}=0$ because she ``truly'' believes the interference
does not exist. Rather, she makes it because it enables her to pin
down the effect of interest from the data. If the researcher could
somehow be informed of the true value of the interference fixed parameter
$\omega_{2}$ (say, by a ``dry run'' that calibrates the measurement
instrument), she would not impose the assumption $\theta^{*}=0$ because
this would not be required to identify $\omega_{1}$.

The models also differ in how predictably the researcher's subjective
beliefs evolve. Correctly specified Bayesian learning satisfies the
Martingale property: the expectation of $\mu_{t+1}\left(E\right)$
equals to $\mu_{t}\left(E\right)$ for all events $E$ conditional
on any history. Under misspecified Bayesian learning, this property
generally fails. A Bayesian outsider with the same prior as the researcher
who observes and correctly utilizes all information expects the researcher's
beliefs to violate it. However, a misspecified Bayesian learner expects
her own beliefs to satisfy the Martingale property because she does
not question whether her model is misspecified. In contrast, an assumption-based
learner would expect to violate the property. In other words, both
the objective and subjective expectations over the researcher's posterior
belief depart from her prior. If she were to form a belief at period
$t$ regarding her beliefs at period $t+1$, she would use the true
context $\theta^{t}$ to calculate the distribution over $\mu_{t+1}\left(E\right)$.
Yet, the posteriors themselves are arrived at by updating $\mu_{t}$
as if $\theta^{*}=0$. 

\section{An Extension: Multiple and ``Structural'' Assumptions\label{sec:Multiple/Structural}}

Our model is restrictive in several respects. First, it assumes a
single feasible identifying assumption rather than a set of identification
strategies that the researcher can choose from. Second, it assumes
the researcher has a single question, rather than a set of questions.
Third, it focuses on assumptions about context parameters rather than
more ``structural'' assumptions about the fixed parameters. In this
section we present two examples that go beyond these restrictions
and offer stylized representations of familiar identification methods
in empirical economics. In both examples, we use KL divergence for
the plausibility criterion.

\subsection{Learning by ``Calibration''\label{subsec:Learning-by-Calibration}}

Suppose that the statistic follows the process
\[
s^{t}=\left(\omega_{1}+\omega_{2}\right)+\varepsilon^{t}
\]
where $\varepsilon^{t}\sim N\left(0,1\right)$. The researcher wants
to learn both $\omega_{1}$ and $\omega_{2}$, i.e., $Q=\{1,2\}$.
There are no context parameters in this specification, hence our notion
of identifying assumptions in the basic model is moot. Clearly, the
researcher cannot identify both fixed parameters from observations
of $s$. However, the researcher can settle for identification of
one of the fixed parameters, by imposing an assumption about the value
of the other fixed parameter. This is an example of an identification
strategy which does not aim at a complete answer to the research question
and settles for a partial answer instead.

Formally, assume that at every period, the researcher can assume $\omega_{2}=\omega_{2}^{*}$
or $\omega_{1}=\omega_{1}^{*}$, where $\omega_{2}^{*}$ and $\omega_{1}^{*}$
can take any value. When the researcher assumes $\omega_{i}=\omega_{i}^{*}$,
she interprets all the variation in $s^{t}$ as a consequence of $\omega_{-i}$
and the sampling error $\varepsilon^{t}$. When the researcher assumes
$\omega_{i}=\omega_{i}^{*}$, she updates only her belief about $\omega_{-i}$.
The researcher selects the assumption that minimizes KL divergence
(relative to the true data-generating process, given her current beliefs),
and performs the research only if this minimal divergence does not
exceed $K>0$.

This learning process is a metaphor for the ``calibration'' method
employed by quantitative macroeconomists. In this field, it is customary
to confront a multi-parameter model with observational data lacking
the richness that enables full identification of the model's parameters.
Macroeconomists then proceed by assigning values to some of the parameters
in order to identify the remaining parameters from the data.

We examine the learning dynamics that this procedure induces. Suppose
that entering period $t$, the researcher's believes that $\omega_{i}\sim N\left(m_{i}^{t},(\sigma_{i}^{t})^{2}\right)$,
independently across the components of $\omega$. Then, 
\[
2D_{KL}\left(p_{S}(\cdot|h^{t})||p_{S}(\cdot|h^{t},\omega_{i}=\omega_{i}^{*})\right)=\frac{\left((\sigma_{1}^{t})^{2}+(\sigma_{2}^{t})^{2}\right)+\left(m_{i}^{t}-\omega_{i}^{*}\right)^{2}}{(\sigma_{-i}^{t})^{2}}-\ln\left(\frac{(\sigma_{1}^{t})^{2}+(\sigma_{2}^{t})^{2}}{(\sigma_{-i}^{t})^{2}}\right)-1.
\]
The divergence minimizing value of $\omega_{i}^{*}$ is $\omega_{i}^{*}=m_{i}^{t}$,
and then 
\[
2D_{KL}\left(p_{S}(\cdot|h_{t})||p_{S}(\cdot|h_{t},\omega_{i}=m_{i}^{t})\right)=\left(\frac{\sigma_{i}^{t}}{\sigma_{-i}^{t}}\right)^{2}-\ln\left(1+\left(\frac{\sigma_{i}^{t}}{\sigma_{-i}^{t}}\right)^{2}\right).
\]
The researcher effectively chooses between setting $\omega_{1}=m_{1}^{t}$
and setting $\omega_{2}=m_{2}^{t}$. The former induces a lower divergence
than the latter if and only if $\sigma_{1}^{t}<\sigma_{2}^{t}$. Therefore,
the researcher will assume there is no uncertainty about the parameter
she is more certain about. This again brings to mind the ``calibration''
methodology: the researcher ``calibrates'' the parameter she is
more confident about, using her best estimate of this parameter.

We assume that $K$ is large enough such that learning always take
place, and that $\sigma_{1}^{1}=\sigma_{2}^{1}$. At $t=1$, the researcher
assumes $\omega_{2}=m_{2}^{1}$, and then updates her belief about
$\omega_{1}$. Because the researcher's belief about each parameter
is given by an independent Gaussian distribution, each update about
$\omega_{i}$ shrinks $\sigma_{i}$ by a deterministic percentage.
In the second period, her beliefs about $\omega_{2}$ have higher
variance, and so she assumes that $\omega_{1}=m_{1}^{t}$ and updates
her beliefs about $\omega_{2}$. She proceeds by assuming that $\omega_{2}=m_{2}^{t}$
and updating her beliefs about $\omega_{1}$ in odd periods and assuming
that $\omega_{1}=m_{1}^{t}$ and updating her beliefs about $\omega_{2}$
in even periods.
\begin{prop}
\label{prop:calibration}As $t\rightarrow\infty$, $\sigma_{i}^{t}\rightarrow0$
almost surely for each $i$. Conditional on the realized value of
$(\omega_{1},\omega_{2})$, $m_{1}^{t}+m_{2}^{t}\rightarrow\omega_{1}+\omega_{2}$
with probability one, and there exists $v>0$ such that $m_{i}^{t}$
is almost surely normally distributed with variance greater than v
for all $t$.
\end{prop}
In the long run, the researcher correctly learns the sum of the two
fixed parameters. She becomes perfectly confident of her estimates
of the individual parameters. However, because $m_{i}^{t}$ has positive
variance conditional the true state for all $t$, her beliefs about
each $\omega_{i}$ are noisy and incorrect with probability one. In
contrast, the usual law of large numbers would imply that the variance
of $m_{i}^{t}$ would go to zero. The learning process also exhibits
order effects. Early observations effectively get more weight than
late ones, and they have a non-vanishing contribution to the limit
belief.

A Bayesian learner in this setting would also have correct beliefs
about $\omega_{1}+\omega_{2}$. In contrast, she would not be certain
about either $\omega_{1}$ or $\omega_{2}$: There would exist $k>0$
such that $\sigma_{i}^{t}>0$ for each $i$. Also, there would be
no order effects. Any permutation of the order in which statistics
arrive leads to the same beliefs.

The result generalizes to unequal variances as follows. Labeling so
that $\sigma_{1}^{1}>\sigma_{2}^{1}$, the research begins by assuming
that $\omega_{2}=m_{2}^{1}$. After updating about $\omega_{1}$ for
some number $k$ of periods, the variance of $\sigma_{1}^{t+k}$ falls
below that of $\sigma_{2}^{t+k}$. At that point, the researcher switches
to the other assumption, namely $\omega_{1}=m_{1}^{t+k}$ and proceeds
to update about $\omega_{2}$. She then repeats, alternating between
updating about $\omega_{1}$ and $\omega_{2}$. A similar conclusion
to Proposition \ref{prop:calibration} obtains.

\subsection{A ``Heckman'' Selection Model\label{subsec:A-Heckman-Selection}}

We now consider the classic problem of drawing causal inferences from
a selective sample. An observational dataset has three variables.
The first ($s_{1}$) indicates whether an agent enters some market
($s_{1}=1$ means entry). It also contains the agent's income conditional
on entry ($s_{3}$) and an exogenous variable ($s_{2}$) that may
affect both the entry decision and the income conditional on entry.
Formally, there are three statistics, $s_{1}$, $s_{2}$ and $s_{3}$,
where $s_{1},s_{2}\in\{0,1\}$ and $s_{3}\in\mathbb{R}.$ Data about
income is available only for agents who enter the market.

Our researcher has two feasible identification strategies to deal
with selection. First, she can make a contextual assumption that market
entry is purely random, thus assuming away selective entry. Second,
she can make an assumption about the fixed parameters in the manner
of ``Heckman correction'' \citep{heckman1979}. We explore the trade-off
between the two methods and how it affects research dynamics.

Formally, the true data-generating process is given by the following
equations. First, $s_{2}^{t}$ is uniformly and independently distributed
over $\{0,1\}$. Second,
\[
s_{1}^{t}=\begin{cases}
\mathbb{I}_{+}\left(s_{2}^{t}+u^{t}\right) & \text{with probability }\theta^{t}\\
\mathbb{I}_{+}\left(s_{2}^{t}+\varepsilon_{1}^{t}\right) & \text{with probability }1-\theta^{t}
\end{cases}
\]
where $\mathbb{I}_{+}(x)=1$ if $x\geq0$ and otherwise equals $0$.
Finally, given $s_{1}^{t}=1$,
\[
s_{3}^{t}=\omega_{1}+\omega_{2}s_{2}^{t}+\omega_{3}\mathbb{E}\left[u^{t}|s_{1}^{t}=1,s_{2}^{t},\theta^{t}\right]+\varepsilon_{2}^{t}
\]
where $u^{t}$, $\varepsilon_{1}^{t}$ and $\varepsilon_{2}^{t}$
are all independent normal variables with mean zero, and where the
variances of $u^{t}$ and $\varepsilon_{1}^{t}$ are the same. The
statistic $s_{3}^{t}$ is not measured when $s_{1}^{t}=0$. The context
parameter $\theta^{t}\in[0,1]$ indicates the probability that an
agent's assignment into the market is based on the agents' latent
characteristics. Thus, $\theta^{t}=0$ means purely random, non-selective
assignment.

There are three fixed parameters in this specification, all of which
enter the equation for $s_{3}^{t}$. These parameters represent the
causal effects of three factors on agents' income: market entry itself
($\omega_{1}$), the exogenous variable $s_{2}^{t}$ ($\omega_{2})$,
and the latent variable $u^{t}$ ($\omega_{3}$). The researcher is
interested in learning $\omega_{1}$, i.e., $Q=\{1\}$. Long-run observation
of $s_{3}^{t}$ for each $s_{2}^{t}$ provides two equations with
three unknowns, hence $\omega_{1}$ cannot be identified unless the
researcher imposes an assumption. Parameterize beliefs $\mu$ so that
$\omega_{i}\sim N\left(m_{i},\sigma_{i}^{2}\right)$.

There are two feasible identifying assumptions. One assumption is
$\theta^{*}=0$, i.e., market entry is independent of $u^{t}$ in
the dataset. Under this assumption, $\mathbb{E}\left[u^{t}|s_{1}^{t}=1,s_{2}^{t}\right]=0$
for every $s_{2}^{t}$, and the long-run average of $s_{3}^{t}$ given
$s_{1}^{t}=1$ and $s_{2}^{t}$ is equal to $\omega_{1}+\omega_{2}s_{2}^{t}$.
This gives two equations with two unknowns and enables the researcher
to pin down $\omega_{1}$.

Alternatively, the researcher could assume that the fixed parameter
$\omega_{2}$ equals $\omega_{2}^{*}=0$. That is, the researcher
assumes that in the dataset, the exogenous variable that may affect
market entry does not have a direct causal effect on income conditional
on entry. It is an ``exclusion'' restriction that turns $s_{2}^{t}$
into a valid instrument for estimating $\omega_{1}$, albeit with
different parameterization than in the IV example we examined in Section
\ref{sec:Limiting Beliefs}.

The second identification method is based on Heckman's correction
method \citep{heckman1979}. For the sake of tractability, we simplified
the model by admitting no fixed parameters into the distribution of
$s_{1}^{t}$ conditional on $s_{2}^{t}$. This enables us to treat
$\mathbb{E}\left[u^{t}|s_{1}^{t}=1,s_{2}^{t}\right]$ as a known quantity,
whereas in practice it would be an estimated one. Our example thus
trivializes the first stage of Heckman's procedure, and focuses on
the second stage.

In any given period, the researcher selects the KL divergence minimizing
assumption ($\theta^{*}=0$ or $\omega_{2}^{*}=0$), as long as this
divergence does not exceed the constant $K$. The following result
characterizes the researcher's selection strategy.
\begin{prop}
\label{prop: Heckman}For almost every history $h^{t}$, there exist
thresholds $0<\bar{\theta}^{RD}\left(\mu\left(h^{t}\right)\right)\leq\bar{\theta}^{S}\left(\mu\left(h^{t}\right)\right)\leq1$
such that the researcher assumes $\theta^{*}=0$ when $\theta^{t}\in\left[0,\bar{\theta}^{RD}\left(\mu\left(h^{t}\right)\right)\right)$;
assumes $\omega_{2}^{*}=0$ when $\theta^{t}\in\left(\bar{\theta}^{S}\left(\mu\left(h^{t}\right)\right),1\right)$;
and passes when $\theta^{t}\in\left(\bar{\theta}^{RD}\left(\mu\left(h^{t}\right)\right),\bar{\theta}^{S}\left(\mu\left(h^{t}\right)\right)\right)$.
The thresholds $\bar{\theta}^{RD}\left(\mu\left(h^{t}\right)\right)$
and $\bar{\theta}^{S}\left(\mu\left(h^{t}\right)\right)$ increase
in $\left(\mathbb{\mathbb{E}}_{\mu\left(h^{t}\right)}(\omega_{2})\right)^{2}$
and $Var_{\mu\left(h^{t}\right)}(\omega_{2})$, and decrease in $\left(\mathbb{\mathbb{E}}_{\mu\left(h^{t}\right)}(\omega_{3})\right)^{2}$.
If $K$ is large enough, then $\bar{\theta}^{RD}\left(\mu\left(h^{t}\right)\right)=\bar{\theta}^{S}\left(\mu\left(h^{t}\right)\right)$.
\end{prop}
Thus, when market entry exhibits little selection (i.e., $\theta$
is small), the researcher employs the contextual assumption $\theta^{*}=0$.
In contrast, when entry is highly selective, the researcher passes
or imposes the assumption $\omega_{2}^{*}=0$. Her willingness to
impose the latter assumption increases with its perceived accuracy
(i.e., as $\mathbb{E}(\omega_{2})$ gets closer to zero) and with
her confidence of her estimate --- i.e., as the variance of her belief
over $\omega_{2}$ goes down. Finally, the researcher is less likely
to employ the contextual assumption when she believes that selective
entry has a large effect on income (i.e., when $\mathbb{E}(\omega_{3})$
is far from zero).

This setting has self-reinforcing learning dynamics. The researcher
never updates her beliefs about $\omega_{3}$ when she assumes $\theta^{*}=0$.
Likewise, she never updates her beliefs about $\omega_{2}$ when she
assumes $\omega_{2}^{*}=0$. When she is confident that $\omega_{2}$
is close to zero, she usually assumes $\omega_{2}^{*}=0$ and rarely
updates her belief over $\omega_{2}.$ Therefore, if this belief is
inaccurate, it will take a long time to correct it. Moreover, when
the researcher assumes $\omega_{2}^{*}=0$, she misattributes part
of the actual effect of $\omega_{2}$ on income to $\omega_{3}$.
Depending on the true values of these parameters, this misattribution
can make the researcher even less likely to employs the contextual
assumption. Similarly, if the researcher is confident that $\omega_{3}$
is low, she tends to assume $\theta^{*}=0$. This leads her to misattribute
part of the actual effect of $\omega_{3}$ on income to $\omega_{2}$,
which may further strengthen her tendency to employ the contextual
assumption. Thus, the researcher's predilection to stick to a particular
identifying strategy for a long stretch of time is history-dependent.

\appendix

\section{Proofs}

For the proofs of Propositions \ref{prop: non-expansion}, \ref{prop: quasi-convex},
and \ref{prop: long run beliefs}, we economize on notation by taking
a history $h^{t}$ and writing $\left(h^{t},s\right)$ for the history
that concatenates $h^{t}$ with the tuple $\left(s^{t}=s,a^{t}=1,\theta^{t}\right)$
for arbitrary $\theta^{t}\in\Theta^{R}\left(\mu\left(h^{t}\right)\right)$
(similarly for $\left(h^{t},s,s^{\prime},s^{\prime\prime},...\right)$).
In addition, for any $\theta,h,s$, let $q(\theta,h)$ and $q(\theta,h,s)$
serve as shorthand notation for the conditional distributions $p_{S,U}\left(\cdot|\theta,h\right)$
and $p_{S,U}\left(\cdot|\theta,\left(h,s\right)\right)$, respectively.

\subsection{Proof of Proposition \ref{prop: non-expansion}}

We begin the proof by drawing a simple implication from the fact that
f-divergences are convex (see page 56 in \citet{amari2000methods};
it follows from arguments similar to Theorem 2.7.2 of \citet{CoverThomas2006}).

The researcher always updates her belief as if the assumption $\theta^{*}$
is correct. Therefore, her belief over variables satisfies the Martingale
property with respect to the distribution over statistics conditional
on the assumption. That is, for every history $h^{\tau}$, context
$\theta^{\tau}$, and statistic realization $s^{\tau}$,
\begin{equation}
\sum_{s^{\tau}\in S}p\left(s^{\tau+1},u^{\tau+1}|\theta^{\tau+1},\left(h^{\tau},s^{\tau}\right)\right)p\left(s^{\tau}|\theta^{*},h^{\tau}\right)=p\left(s^{\tau+1},u^{\tau+1}|\theta^{\tau+1},h^{\tau}\right).\label{eq:martingale}
\end{equation}
For completeness, we derive this equation in greater detail. The researcher's
belief over the variable realizations at $\tau+1$ given $\theta^{\tau+1},\left(h^{\tau},s^{\tau}\right)$
is 
\[
p\left(s^{\tau+1},u^{\tau+1}|\theta^{\tau+1},\left(h^{\tau},s^{\tau}\right)\right)=\int_{\omega}p\left(s^{\tau+1},u^{\tau+1}|\theta^{\tau+1},\omega\right)\mu\left(\omega|h^{\tau},s^{\tau}\right)d\omega.
\]
Since the researcher always updates her beliefs as if $\theta=\theta^{*}$,
as given by Eq (\ref{eq:belief updating}),
\[
p\left(s^{\tau+1},u^{\tau+1}|\theta^{\tau+1},\left(h^{\tau},s^{\tau}\right)\right)=\int_{\omega}p\left(s^{\tau+1},u^{\tau+1}|\theta^{\tau+1},\omega\right)\frac{p\left(s^{\tau}|\theta^{*},\omega\right)}{p\left(s^{\tau}|\theta^{*},h^{\tau}\right)}\mu\left(\omega|h^{\tau}\right)d\omega.
\]
Multiplying by $p\left(s^{\tau}|\theta^{*},h^{\tau}\right)$ and summing
across $S$, we obtain
\begin{align*}
 & \sum_{s^{\tau}\in S}p\left(s^{\tau+1},u^{\tau+1}|\theta^{\tau+1},\left(h^{\tau},s^{\tau}\right)\right)p\left(s^{\tau}|\theta^{*},h^{\tau}\right)\\
 & =\sum_{s^{\tau}\in S}\int_{\omega}p\left(s^{\tau+1},u^{\tau+1}|\theta^{\tau+1},\omega\right)p\left(s^{\tau}|\theta^{*},\omega\right)\mu\left(\omega|h^{\tau}\right)d\omega\\
 & =\int_{\omega}p\left(s^{\tau+1},u^{\tau+1}|\theta^{\tau+1},\omega\right)\sum_{s^{\tau}\in S}p\left(s^{\tau}|\theta^{*},\omega\right)\mu\left(\omega|h^{\tau}\right)d\omega\\
 & =\int_{\omega}p\left(s^{\tau+1},u^{\tau+1}|\theta^{\tau+1},\omega\right)\mu\left(\omega|h^{\tau}\right)d\omega=p\left(s^{\tau+1},u^{\tau+1}|\theta^{\tau+1},h^{\tau}\right).
\end{align*}
By Eq (\ref{eq:martingale}) and the convexity of f-divergence,
\[
D\left(q(\theta^{\tau},h^{\tau})||q(\theta^{*},h^{\tau})\right)\leq\sum_{s^{\tau}\in S}D\left(q(\theta^{\tau},h^{\tau},s^{\tau})||q(\theta^{*},h^{\tau},s^{\tau})\right)p\left(s^{\tau}|\theta^{*},h^{\tau}\right).
\]

Our next step is to show that if the divergence drops after some statistic
realization, this implies a lower bound on its increase under some
alternative realization. For any history $h$ and statistic realization
$s$, let
\[
\delta\left(\theta,h,s\right)\equiv D\left(q(\theta,h)||q(\theta^{*},h)\right)-D\left(q(\theta,h,s)||q(\theta^{*},h,s)\right)
\]
 be the change in the divergence given $\theta$ when the history
$h$ is concatenated by $s$. 
\begin{claim}
If $\delta\left(\theta,h,s\right)>0$, then there exists $s'$ such
that 
\[
D\left(q(\theta,h,s')||q(\theta^{*},h,s')\right)\geq D\left(q(\theta,h)||q(\theta^{*},h)\right)+p(s|\theta^{*},h)\delta\left(\theta,h,s\right).
\]
\begin{proof}
Denote $q(\theta,h,-s)=p_{S,U}\left(\cdot|\theta,\left\{ \left(h,s^{\prime}\right)\right\} _{s^{\prime}\neq s}\right).$
By convexity of f-divergence,
\begin{align*}
 & D\left(q(\theta,h)||q(\theta^{*},h)\right)\\
\leq & p(s|\theta^{*},h)D\left(q(\theta,h,s)||q(\theta^{*},h,s)\right)+\left(1-p(s|\theta^{*},h)\right)D\left(q(\theta,h,-s)||q(\theta^{*},h,-s)\right)
\end{align*}
This inequality can be rewritten as
\begin{align*}
 & D\left(q(\theta,h,-s)||q(\theta^{*},h,-s)\right)-D\left(q(\theta,h)||q(\theta^{*},h)\right)\\
\geq & p(s|\theta^{*},h)\left[D\left(q(\theta,h,-s)||q(\theta^{*},h,-s)\right)-D\left(q(\theta,h,s)||q(\theta^{*},h,s)\right)\right].
\end{align*}
Moreover, since $\delta\left(\theta,h,s\right)>0$, the R.H.S. of
this inequality is weakly above
\[
p(s|\theta^{*},h)\left[D\left(q(\theta,h)||q(\theta^{*},h)\right)-D\left(q(\theta,h,s)||q(\theta^{*},h,s)\right)\right]=p(s|\theta^{*},h)\delta\left(\theta,h,s\right).
\]
By convexity of f-divergence and the definition of $q(\theta,h,-s)$,
there is $s'\neq s$ such that
\[
D\left(q(\theta,h,s')||q(\theta^{*},h,s')\right)\geq D\left(q(\theta,h,-s)||q(\theta^{*},h,-s)\right),
\]
which completes the proof.
\end{proof}
\end{claim}
Fix a history $h^{t+1}$ and $\theta^{t+1}$ so that $\theta^{t+1}\in\varTheta^{R}\left(\mu\left(h^{t+1}\right)\right)\setminus\varTheta^{R}\left(\mu\left(h^{t}\right)\right)$.
Consider any $h^{t}$ and $s^{*}$ such that 
\[
D\left(q\left(\theta,h^{t},s^{*}\right)||q\left(\theta^{*},h^{t},s^{*}\right)\right)\leq K<D\left(q\left(\theta,h^{t}\right)||q\left(\theta^{*},h^{t}\right)\right).
\]
Assume the proposition is false. Then, then for any $T\geq t+2$ and
any continuation of statistic realizations $\left(s^{t+2},s^{t+3},\dots,s^{T}\right)$,
we have
\[
D\left(q\left(|\theta,\left(h^{t},s^{*},s^{t+2},s^{t+3},\dots,s^{T}\right)\right)||q\left(\theta^{*},\left(h^{t},s^{*},s^{t+2},s^{t+3},\dots,s^{T}\right)\right)\right)\leq K.
\]
Since Bayesian posterior beliefs are invariant to the ordering of
a sequence of conditionally independent signals, this inequality can
be equivalently rewritten as
\begin{equation}
D\left(q\left(\theta,\left(h^{t},s^{t+2},s^{t+3},\dots,s^{T},s^{*}\right)\right)||q\left(\theta^{*},\left(h^{t},s^{t+2},s^{t+3},\dots,s^{T},s^{*}\right)\right)\right)\leq K.\label{eq:D<K}
\end{equation}

We inductively construct a sequence of histories $\bar{h}^{T}$ for
$T\geq t$, starting with $\bar{h}^{t}=h^{t}$, whose divergence goes
to infinity. Inductively assume that 
\begin{equation}
D\left(q\left(\theta,\bar{h}^{T}\right)||q\left(\theta^{*},\bar{h}^{T}\right)\right)\geq D\left(q\left(\theta,h^{t}\right)||q\left(\theta^{*},h^{t}\right)\right)>K,\label{eq:IH prop 1}
\end{equation}
which is true by assumption for $T=t$. Then, by Equation (\ref{eq:D<K}),
\[
D\left(q\left(\theta,\left(\bar{h}^{T},s^{*}\right)\right)||q\left(\theta^{*},\left(\bar{h}^{T},s^{*}\right)\right)\right)\leq K
\]
By the claim, we can find $s^{T+1}$ such that 
\[
D\left(q\left(\theta,\left(\bar{h}^{T},s^{T+1}\right)\right)||q\left(\theta^{*},\left(\bar{h}^{T},s^{T+1}\right)\right)\right)\geq D\left(q\left(\theta,\bar{h}^{T}\right)||q\left(\theta^{*},\bar{h}^{T}\right)\right)+p\left(s^{*}|\theta^{*},\bar{h}^{T}\right)\delta\left(\theta,\bar{h}^{T},s^{*}\right).
\]
Moreover, by Equation (\ref{eq:IH prop 1}),
\begin{equation}
\delta\left(\bar{h}^{T},s^{*}\right)\geq D\left(q\left(\theta,h^{t}\right)||q\left(\theta^{*},h^{t}\right)\right)-K\equiv\delta^{*}>0\label{eq:delta prop 1}
\end{equation}
and 
\[
p\left(s^{*}|\theta^{*},\bar{h}^{T}\right)\geq\min_{\omega}p\left(s^{*}|\theta^{*},\omega\right)\equiv p^{*}>0.
\]
Set $\bar{h}^{T+1}=\left(\bar{h}^{T},s^{T+1}\right)$. Applying the
inductive step, we conclude that
\[
D\left(q\left(\theta,\bar{h}^{T+1}\right)||q\left(\theta^{*},\bar{h}^{T+1}\right)\right)\geq D\left(q\left(\theta,\bar{h}^{T}\right)||q\left(\theta^{*},\bar{h}^{T}\right)\right)+p^{*}\delta^{*}>K.
\]
This means that as $T\rightarrow\infty$,
\[
D\left(p_{S,U}\left(\cdot|\theta,\bar{h}^{T}\right)||p_{S,U}\left(\cdot|\theta^{*},\bar{h}^{T}\right)\right)\rightarrow\infty.
\]
However, by convexity of $D$, the L.H.S. is bounded from above by
\[
\max_{\omega\in\Omega}D\left(p_{S,U}\left(\cdot|\theta,\omega\right)||p_{S,U}\left(\cdot|\theta^{*},\omega\right)\right)\leq\infty,
\]
a contradiction.$\oblong$

\subsection{Proof of Proposition \ref{prop: quasi-convex}}

Fix any $h^{t}$ and $s$ so that $\Theta^{R}\left(\mu\left(h^{t},s\right)\right)\setminus\Theta^{R}\left(\mu\left(h^{t}\right)\right)\neq\emptyset$.
Pick $\theta^{1}\in\Theta^{R}\left(\mu\left(h^{t},s\right)\right)\setminus\Theta^{R}\left(\mu\left(h^{t}\right)\right)$
for which 
\[
D\left(q\left(\theta^{1},h^{t}\right)||q\left(\theta^{*},h^{t}\right)\right)>K>D\left(q\left(\theta^{1},h^{t},s\right)||q\left(\theta^{*},h^{t},s\right)\right)=\Delta.
\]
By continuity, there exists $\theta=\beta\theta^{1}+(1-\beta)\theta^{*}$
so that 
\[
D\left(q\left(\theta,h^{t}\right)||q\left(\theta^{*},h^{t}\right)\right)=K,
\]
and so $\theta\in\Theta^{R}\left(\mu(h^{t})\right)$. By quasi-convexity,
$D\left(q\left(\theta,h^{t},s\right)||q\left(\theta^{*},h^{t},s\right)\right)<K$.
By convexity of f-divergence (page 56, \citet{amari2000methods}),
\begin{align*}
K & =D\left(q\left(\theta,h^{t}\right)||q\left(\theta^{*},h^{t}\right)\right)\\
 & \leq\sum_{s'}D\left(q\left(\theta,h^{t},s'\right)||q\left(\theta^{*},h^{t},s'\right)\right)p(s'|\theta^{*},h^{t})\\
 & <\sum_{s'\neq s}D\left(q\left(\theta,h^{t},s'\right)||q\left(\theta^{*},h^{t},s'\right)\right)p(s'|\theta^{*},h^{t})+\Delta p(s|\theta^{*},h^{t})
\end{align*}
Therefore, $D\left(q\left(\theta,h^{t},s''\right)||q\left(\theta^{*},h^{t},s''\right)\right)>K$
for some $s''\in S\setminus\{s\}$. This also holds for all $\theta^{\prime}$
sufficiently close to $\theta$, including some of those for which
$D\left(q\left(\theta^{\prime},h^{t}\right)||q\left(\theta^{*},h^{t}\right)\right)<K$.
Therefore, $\Theta^{R}\left(\mu\left(h^{t}\right)\right)\setminus\Theta^{R}\left(\mu\left(h^{t+1}\right)\right)\neq\emptyset$
with a probability of at least $p\left(s'',\Theta^{R}\left(\mu\left(h^{t}\right)\right)|\omega\right)$,
for any given value $\omega$ of the fixed parameters and any history
$h^{t}$. $\oblong$

\subsection{Proof of Proposition \ref{prop: constant rage-1}}

Let the DAG $G=(N,R)$ be the structure of $p$. We introduce a few
pieces of DAG-based notation. First, just as $N^{s}$ is the set of
nodes that represent statistics, $N^{\omega}$ is the set of nodes
that represent fixed parameters. Define $N^{u}$ and $N^{\theta}$
in the same manner. Recall that by definition, the nodes in $N^{\omega},$
$N^{\theta}$ and $N^{u}$ are ancestral. For convenience, we will
sometimes abuse notation and identify $\omega_{i}$, $s_{i}$, $u_{i}$,
and $\theta_{i}$ with their corresponding nodes. The proof proceeds
step-wise.

\textbf{Step 1}: The researcher never learns anything about $\omega_{-Q^{*}}$.

By definition of $Q^{*}$, $p(s|\theta^{*},\omega_{Q^{*}},\omega_{-Q^{*}})=p(s|\theta^{*},\omega_{Q^{*}})$
for every $s$, so for almost every $\omega$,
\begin{align*}
\mu\left(\omega_{-Q^{*}}|h^{t},s^{t},a^{t}=1,\theta^{t}\right) & =\frac{\int p\left(s^{t}|\theta^{*},\omega\right)\mu\left(\omega_{Q^{*}}|h^{t}\right)\mu\left(\omega_{-Q^{*}}|h^{t}\right)d\omega_{Q^{*}}}{\int p\left(s^{t}|\theta^{*},\omega^{\prime}\right)\mu\left(\omega^{\prime}|h^{t}\right)d\omega^{\prime}}\\
 & =\frac{\int p\left(s^{t}|\theta^{*},\omega\right)\mu\left(\omega_{Q^{*}}|h^{t}\right)d\omega_{Q^{*}}}{\int p\left(s^{t}|\theta^{*},\omega_{Q}^{\prime}\right)\mu\left(\omega_{Q^{*}}^{\prime}|h^{t}\right)d\omega_{Q^{*}}^{\prime}}\mu\left(\omega_{-Q^{*}}|h^{t}\right)=\mu\left(\omega_{-Q^{*}}|h^{t}\right)
\end{align*}
Therefore, beliefs about almost every $\omega_{-Q^{*}}$ are history-independent.

In preparation for the next step, define a subset $I\subseteq N^{\omega}$
consisting of all the parameters that are not independent of $\theta$
conditional on $(s,u)$ in the following recursive manner. First,
$I_{0}$ is the set of nodes $i\in N^{\omega}$ for which there exist
$j\in N^{\theta}$ and $k\in N^{s}$ such that $i,j\in R(k)$. For
every integer $n>0$, $I_{n}$ is the set of nodes $i\in N^{\omega}$
for which there exist $j\in I_{n-1}$ and $k\in N^{s}$ such that
$i,j\in R(k)$. Define $I=\cup_{n\geq0}I_{n}$. Let $N^{I}$ be the
nodes in $N^{s}$ with a parent in $I$. By construction, $j\in N^{I}$
implies that $R(j)\cap N^{\omega}\subset I$, whereas $j\notin N^{I}$
implies that $R(j)\cap I=\emptyset$.

\textbf{Step 2}: $I\cap Q^{*}=\emptyset$.

For contradiction, suppose that $\omega_{i}\in I\cap Q^{*}$. Then,
there is a sequence $\omega_{i_{1}},\dots,\omega_{i_{n}}$ of structural-parameter
nodes and a sequence $s_{i_{1}},\dots,s_{i_{n}}$ of statistics nodes,
such that: $\omega_{i}=\omega_{i_{1}}$; every node $s_{i_{k}}$ along
the sequence ($k=1,...,n-1$) is a child of $\omega_{i_{k}}$ and
$\omega_{i_{k+1}}$; and $s_{i_{n}}$ is a child of a node in $N^{\theta}$.
The following diagram illustrates this configuration for $n=3$.
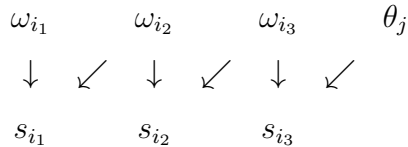
\[
\begin{array}{ccccccc}
\omega_{i_{1}} &  & \omega_{i_{2}} &  & \omega_{i_{3}} &  & \theta_{j}\\
\downarrow & \swarrow & \downarrow & \swarrow & \downarrow & \swarrow\\
s_{i_{1}} &  & s_{i_{2}} &  & s_{i_{3}}
\end{array}
\]
We show that $G$ does not satisfy the conditional-independence property
$s_{i_{1}}\perp\theta|\left(s_{-i_{1}},u\right)$. By a basic result
in the Bayesian-network literature (e.g., \citet{kollerFriedman2009}),
this property has a simple graphical characterization, known as d-separation.
Perform the following two-step procedure.\footnote{In general, there is a preliminary step, in which all nodes that appear
below the nodes that represent $\omega_{i},\theta,s,u$ are removed.
Since there are no such nodes in $G$, this step is vacuous.} First, take every triple of nodes $i,j,k$ such that $i,j\in R(k)$
whereas $i$ and $j$ are not linked. Modify the DAG by connecting
$i$ and $j$. Second, remove the directionality of all links in the
modified graph, such that it becomes a non-directed graph. In this
so-called ``moral graph'' induced by $G$, check whether every path
between $s^{i_{1}}$ and a node in $N^{\theta}$ contains a node in
$N^{s}\cup N^{u}$. This is not the case, by construction, as the
moral graph contains a path from $s_{i_{1}}$ to $\theta$ that goes
through the nodes $\omega_{i_{1}},\dots,\omega_{i_{n}}$. For illustration,
note that procedure generates the following moral graph from the DAG
above:
\[
\begin{array}{ccccccc}
\omega_{i_{1}} & - & \omega_{i_{2}} & - & \omega_{i_{3}} & - & \theta_{j}\\
| & / & | & / & | & /\\
s_{i_{1}} &  & s_{i_{2}} &  & s_{i_{3}}
\end{array}
\]
It follows that $s_{i_{1}}\not\perp\theta|\left(s_{-i_{1}},u\right)$.
By hypothesis, this implies $s_{i_{1}}\perp\omega_{Q^{*}}$, and hence
$s_{i_{1}}\perp\omega_{i}$ (because $\omega_{i}$ is in $Q^{*}$).
Since $s_{i_{1}}$ is a child of $\omega_{i},$ this property is violated,
a contradiction. Therefore, we conclude that $I$ and $Q^{*}$ are
disjoint.

\textbf{Step 3}: For every $s,u$, the likelihood ratio $p(s,u|\theta^{t},h^{t})/p(s,u|\theta^{*},h^{t})$
is history-independent.

For every $j\in N^{s}$, we use $\left(s,u,\omega,\theta\right)_{R(j)}$
to denote the values of the variables and parameters that are represented
by the nodes in $R(j)$. Then, $p(s,u|\theta^{t},h^{t})=p(u)p(s|\theta^{t},h^{t},u)$
and we can write $p(s^{t}|\theta^{t},h^{t},u^{t})$ equals

\begin{align*}
 & \int\prod_{j\in N^{s}}p\left(s_{j}^{t}|\left(s^{t},u^{t},\theta^{t},\omega\right)_{R(j)}\right)\mu(\omega|h^{t})d\omega\\
= & \int\int\prod_{j\in N^{I}}p\left(s_{j}^{t}|\left(s^{t},u^{t},\theta^{t},\omega\right)_{R(j)}\right)\prod_{j\notin N^{I}}p\left(s_{j}^{t}|\left(s^{t},u^{t},\theta^{t},\omega\right)_{R(j)}\right)\mu(\omega_{-I}|h^{t})\mu(\omega_{I}|h^{t})d\omega_{I}d\omega_{-I}\\
= & \left(\int\prod_{j\in N^{I}}p\left(s_{j}^{t}|\left(s^{t},u^{t},\theta^{t},\omega\right)_{R(j)}\right)\mu(\omega_{I}|h^{t})d\omega_{I}\right)\left(\int\prod_{j\notin N^{I}}p\left(s_{j}^{t}|\left(s^{t},u^{t},\theta^{t},\omega\right)_{R(j)}\right)\mu(\omega_{-I}|h^{t})d\omega_{-I}\right)
\end{align*}
where the second equality follows from the relationship between $N^{I}$
and $I$ we articulated above.

By the definition of $N^{I}$, $p\left(s_{j}^{t}|\left(s^{t},u^{t},\theta^{t},\omega\right)_{R(j)}\right)$
is constant in $\theta^{t}$ for every $j\notin N^{I}$. By Step 2,
$Q^{*}\cap I=\emptyset$. By Step 1, $\mu(\omega_{I}|h^{t})$ is constant
in $h^{t}$. It follows that we can write the likelihood ratio as
\[
\frac{p(s^{t},u^{t}|\theta^{t},h^{t})}{p(s^{t},u^{t}|\theta^{*},h^{t})}=\frac{\int\prod_{j\in N^{I}}p\left(s_{j}^{t}|\left(s^{t},u^{t},\theta^{t},\omega\right)_{R(j)}\right)\mu(\omega_{I})d\omega_{I}}{\int\prod_{j\in N^{I}}p\left(s_{j}^{t}|\left(s^{t},u^{t},\theta^{*},\omega\right)_{R(j)}\right)\mu(\omega_{I})d\omega_{I}}
\]
because the other multiplicative terms in $p(s,u|\theta)$ cancel
out. Therefore, the likelihood ratio is history-independent.

\textbf{Step 4}: Completing the proof

Let $R\left(N^{I}\right)=\bigcup_{j\in N^{I}}R(j)$. Suppose $s_{k}$
is in $N^{I}$. As we saw in the proof of Step 2, $s_{k}$ is not
independent of $\theta$ conditional on $(s_{-k},u)$. By hypothesis,
$s_{k}\perp\omega_{Q^{*}}.$ This means that $s_{k}$ cannot be a
descendant of any node in $\omega_{Q^{*}}$ according to $G$. It
follows that the parents of $s_{k}$ also cannot be descendants of
nodes in $\omega_{Q^{*}}.$ Therefore, for every $s_{j}$ node in
$N^{I}\cup R\left(N^{I}\right)$, $p\left(s_{j}^{t}|\left(s^{t},u^{t},\theta^{*},\omega\right)_{R(j)}\right)$
is constant in $\omega_{Q^{*}}$, and so by Step 1, $p\left(s_{N^{I}\cup R\left(N^{I}\right)}|h^{t},\theta^{*}\right)=p\left(s_{N^{I}\cup R\left(N^{I}\right)}|\theta^{*}\right)$
for every history $h^{t}$.

Note that $D(p_{S,U}(\cdot|h^{t},\theta^{t})||p_{S,U}(\cdot|h^{t},\theta^{*}))$
equals
\begin{align*}
 & \sum_{(s,u)}p(u^{t})p(s^{t}|h^{t},\theta^{*},u)f\left(\frac{p(s^{t},u^{t}|\theta^{t},h^{t})}{p(s^{t},u^{t}|\theta^{*},h^{t})}\right)\\
= & \sum_{(s,u)}p(u^{t})p(s^{t}|h^{t},\theta^{*},u)f\left(\frac{\int\prod_{j\in N^{I}}p\left(s_{j}^{t}|\left(s^{t},u^{t},\theta^{t},\omega\right)_{R(j)}\right)\mu(\omega_{I})d\omega_{I}}{\int\prod_{j\in N^{I}}p\left(s_{j}^{t}|\left(s^{t},u^{t},\theta^{*},\omega\right)_{R(j)}\right)\mu(\omega_{I})d\omega_{I}}\right)
\end{align*}
using the simplified expression for the likelihood ratio that we derived
at the end of the proof of Step 3. The only $s$ variables it involves
are those represented by nodes in $N^{I}\cup R(N^{I})$. Therefore,
the likelihood ratio is independent of $s_{-\left(N^{I}\cup R\left(N^{I}\right)\right)}$.
It follows that for each $u$, when we sum over the values of $s_{-\left(N^{I}\cup R\left(N^{I}\right)\right)}$,
their contributions to the divergence are integrated out, and we can
replace $p(s|h^{t},\theta^{*},u)$ with $p\left(s_{N^{I}\cup R\left(N^{I}\right)}|\theta^{*},u\right)$
in the expression above. We have already observed that the likelihood
ratio is history-independent for every $s_{N^{I}}$, as is the distribution
of $s_{N^{I}\cup R\left(N^{I}\right)}$. Therefore, the divergence
simplifies into the following history-independent expression
\[
\sum_{(s,u)}p(u^{t})p\left(s_{N^{I}\cup R\left(N^{I}\right)}^{t}|\theta^{*},u\right)f\left(\frac{\int\prod_{j\in N^{I}}p\left(s_{j}^{t}|\left(s^{t},u^{t},\theta^{t},\omega\right)_{R(j)}\right)\mu(\omega_{I})d\omega_{I}}{\int\prod_{j\in N^{I}}p\left(s_{j}^{t}|\left(s^{t},u^{t},\theta^{*},\omega\right)_{R(j)}\right)\mu(\omega_{I})d\omega_{I}}\right),
\]
completing the proof. $\oblong$

\subsection{Proof of Proposition \ref{prop: long run beliefs}}

Consider any stable density $\mu$. There is a positive probability
set of histories $H$ so that $\mu\left(\cdot|h^{t}\right)\rightarrow\mu$
for every $h\in H$. Denote by $\mu_{t+1}(\cdot)$ the Borel measure
with density $\mu\left(\cdot|h^{t}\right)$ and $\mu^{*}$ the Borel
measure with density $\mu$; note $\mu_{t+1}$ is implicitly a function
of $h^{t}$. By the Portmanteau theorem, $\mu_{t+1}$ weak{*} converges
to $\mu^{*}$ for every $h\in H$. Take any open set $O\supset\arg\min_{\omega^{\prime}\in\Omega}D_{KL}\left(p_{S}\left(\cdot|\theta\in\varTheta^{R}\left(\mu^{*}\right),\omega^{*}\right)||p_{S}\left(\cdot|\theta^{*},\omega^{\prime}\right)\right)$
and let $C=\Omega\setminus O$, noting $C$ is compact as a closed
subset of $\Omega$.

Pick any $w\in C$ and any $\hat{w}\in\arg\min_{\omega^{\prime}\in\Omega}D_{KL}\left(p_{S}\left(\cdot|\theta\in\varTheta^{R}\left(\mu^{*}\right),\omega^{*}\right)||p_{S}\left(\cdot|\theta^{*},\omega^{\prime}\right)\right)$.
For any $s\in S$ and states $\omega,\omega^{\prime}\in\Omega$, define
\[
\Delta\left(s,\omega,\omega^{\prime}\right)\equiv p\left(s|\theta^{*},\omega\right)-p\left(s^{t}|\theta^{*},\omega^{\prime}\right).
\]
We show there is an open $E\ni w$ with $\mu^{*}(E)=0$. By continuity
and that $S$ is finite, we can find open sets $\hat{E}\ni\hat{w}$
and $E\ni w$ so that $|\Delta\left(s,\omega,\hat{w}\right)|,|\Delta\left(s,\omega^{\prime},w\right)|<\epsilon$
for all $s\in S$, $\omega\in E$, and $\omega^{\prime}\in\hat{E}$,
and where $\epsilon>0$ satisfies
\[
\sum_{s\in S}p\left(s|\Theta^{R}\left(\mu^{*}\right),\omega^{*}\right)\left[\ln\left(\frac{p\left(s|\theta^{*},\hat{w}\right)-\epsilon}{p\left(s|\theta^{*},w\right)+\epsilon}\right)\right]>0.
\]
Such an $\epsilon$ exists since $\hat{w}$ minimizes divergence,
because then
\begin{align*}
0> & D_{KL}\left(p\left(s|\Theta^{R}\left(\mu^{*}\right),\omega^{*}\right)||p(s|\theta^{*},\hat{w})\right)-D_{KL}\left(p\left(s|\Theta^{R}\left(\mu^{*}\right),\omega^{*}\right)||p(s|\theta^{*},w)\right)\\
= & \sum_{s\in S}p\left(s|\Theta^{R}\left(\mu^{*}\right),\omega^{*}\right)\left[\ln\left(\frac{p\left(s|\theta^{*},w\right)}{p\left(s|\theta^{*},\hat{w}\right)}\right)\right].
\end{align*}

Now, for every history
\begin{align*}
\mu_{t+1}\left(\hat{E}\right) & =p\left(s^{t}|\theta^{*},h^{t}\right)^{-1}\int_{\omega\in\hat{E}}p\left(s^{t}|\theta^{*},\omega\right)d\mu_{t}\\
 & =p\left(s^{t}|\theta^{*},h^{t}\right)^{-1}\int_{\omega\in\hat{E}}\left\{ p\left(s^{t}|\theta^{*},\hat{w}\right)+\Delta\left(s^{t},\omega,\hat{w}\right)\right\} d\mu_{t}
\end{align*}
 and so 
\[
\mu_{t+1}\left(\hat{E}\right)p\left(s^{t}|\theta^{*},h^{t}\right)\mu_{t}\left(\hat{E}\right)^{-1}-p\left(s^{t}|\theta^{*},\hat{w}\right)\in\left(\inf_{s,\omega\in\hat{E}}\Delta\left(s,\omega,\hat{w}\right),\sup_{s,\omega\in\hat{E}}\Delta\left(s,\omega,\hat{w}\right)\right).
\]
Similarly, 
\[
\mu_{t+1}\left(E\right)p\left(s^{t}|\theta^{*},h^{t}\right)\mu_{t}\left(E\right)^{-1}-p\left(s^{t}|\theta^{*},w\right)\in\left(\inf_{s,\omega\in E}\Delta\left(s,\omega,w\right),\sup_{s,\omega\in E}\Delta\left(s,\omega,w\right)\right).
\]
Therefore, for every $t$ there exist $\delta_{t},\hat{\delta}_{t}\in\left(-\epsilon,\epsilon\right)$
so that
\[
\frac{\mu_{t+1}(\hat{E})}{\mu_{t+1}(E)}=\frac{\mu_{t}(\hat{E})}{\mu_{t}(E)}\frac{p\left(s^{t}|\theta^{*},\hat{w}\right)+\hat{\delta}_{t}}{p\left(s^{t}|\theta^{*},w\right)+\delta_{t}}
\]
when $s^{t}$ occurs and $\theta^{t}\in\Theta^{R}\left(\mu\left(h^{t}\right)\right)$.

We claim that $\mu^{*}(E)=0$. Suppose not, so $\mu^{*}(E)>0$ and
so $\mu_{0}(E)>0$. In the history $h^{t}=\left(a^{1},s^{1},\theta^{1};a^{2},s^{2},\theta^{2};\dots;a^{t-1},s^{t-1},\theta^{t-1}\right)$
we have
\begin{align}
\ln\frac{\mu_{t+1}(\hat{E})}{\mu_{t+1}(E)} & =\ln\frac{\mu_{t}(\hat{E})}{\mu_{t}(E)}+\mathbb{I}_{\Theta^{R}\left(\mu\left(h^{t}\right)\right)}(\theta^{t})\ln\frac{p\left(s^{t}|\theta^{*},\hat{w}\right)+\hat{\delta}_{t}}{p\left(s^{t}|\theta^{*},w\right)+\delta_{t}}\label{eq:updating rule-1}\\
 & =\ln\frac{\mu_{0}(\hat{E})}{\mu_{0}(E)}+\sum_{\tau=1}^{t}\mathbb{I}_{\Theta^{R}\left(\mu\left(h^{\tau}\right)\right)}(\theta^{\tau})\ln\frac{p\left(s^{t}|\theta^{*},\hat{w}\right)+\hat{\delta}_{t}}{p\left(s^{t}|\theta^{*},w\right)+\delta_{t}}\nonumber \\
 & \geq\ln\frac{\mu_{0}(\hat{E})}{\mu_{0}(E)}+\sum_{\tau=1}^{t}\mathbb{I}_{\Theta^{R}\left(\mu\left(h^{\tau}\right)\right)}(\theta^{\tau})\ln\frac{p\left(s|\theta^{*},\hat{w}\right)-\epsilon}{p\left(s|\theta^{*},w\right)+\epsilon}\nonumber 
\end{align}
Let 
\[
\bar{l}\left(\mu\right)=\int_{\Theta^{R}\left(\mu\right)}\left[\sum_{s^{\prime}\in S}p\left(s^{\prime}|\theta,\omega^{*}\right)\ln\frac{p\left(s|\theta^{*},\hat{w}\right)-\epsilon}{p\left(s|\theta^{*},w\right)+\epsilon}\right]dp(\theta)
\]
Then,

\begin{align*}
\frac{1}{t}\ln\frac{\mu_{t+1}(\hat{E})}{\mu_{t+1}(E)}\geq & \frac{1}{t}\left[\ln\frac{\mu_{0}(\hat{E})}{\mu_{0}(E)}+\sum_{\tau=1}^{t}\bar{l}\left(\mu_{\tau}\right)\right]\\
 & \qquad+\frac{1}{t}\sum_{\tau=1}^{t}\left[\mathbb{I}_{\Theta^{R}\left(\mu_{\tau}\right)}(\theta^{\tau})\ln\frac{p\left(s|\theta^{*},\hat{w}\right)-\epsilon}{p\left(s|\theta^{*},w\right)+\epsilon}-\bar{l}\left(\mu_{\tau}\right)\right].
\end{align*}
By arguments that are substantively identical to Claim B of \citet{espondaPouzo2016},
\begin{equation}
\frac{1}{t}\sum_{\tau=1}^{t}\left[\mathbb{I}_{\Theta^{R}\left(\mu_{\tau}\right)}(\theta^{\tau})\ln\frac{p\left(s|\theta^{*},\hat{w}\right)-\epsilon}{p\left(s|\theta^{*},w\right)+\epsilon}-\bar{l}\left(\mu_{\tau}\right)\right]\rightarrow0\label{eq:LLN to zero-1}
\end{equation}
almost surely given $\omega^{*}$ and that $h^{\tau}\in H$. It follows
from $\mathbb{P}\left(\mu\left(\cdot|h^{t}\right)\rightarrow\mu^{*}|H,\omega^{*}\right)=1$
and continuity of $\Theta^{R}(\cdot)$ at $\mu^{*}$ that
\[
\mathbb{P}\left(\lim_{t}\left|\bar{l}\left(\mu_{t}\right)-\bar{l}\left(\mu^{*}\right)\right|=0|H,\omega^{*}\right)=1.
\]
Therefore, 
\[
\mathbb{P}\left(\lim_{t}\frac{1}{t}\ln\frac{\mu_{t+1}(\hat{E})}{\mu_{t+1}(E)}\geq\bar{l}(\mu^{*})|H,\omega^{*}\right)=1
\]
and since $\bar{l}(\mu^{*})>0$ by construction, we must have $\lim\frac{1}{t}\ln\frac{\mu_{t+1}(\hat{E})}{\mu_{t+1}(E)}>0$,
which requires $\mu_{t+1}(E)\rightarrow0$, contradicting that $\mu^{*}(E)>0$.

Now, for each $w\in C$, let $E_{w}$ be the open set constructed
above. $\left\{ E_{w}:w\in C\right\} $ is an open cover of $C$ and
so has a finite sub-cover $\left\{ E_{1},\dots,E_{k}\right\} $. The
set $E^{*}=\cup_{i=1}^{k}E_{i}\supset C$, and $\mu^{*}(E^{*})=0$.
Conclude $\mu^{*}(O)=1$. $\oblong$

\subsection{Proof of Proposition \ref{prop: generic incredible certitude}}

Because $D$ is strictly increasing in $\theta$, $\Theta^{R}(\mu^{\prime})=[0,\bar{\theta}(\mu^{\prime})]$
for some $\bar{\theta}:\Delta\Omega\rightarrow(0,1]$. Since $D$
is continuous, $\bar{\theta}$ is continuous in $\mu$ by the Berge
Maximum Theorem, and thus $\Theta^{R}$ is a continuous correspondence.
Observe that by definition of $Q^{*}$, (\ref{eq:KL for stable})
does not depend on $\omega_{-Q^{*}}^{\prime}$. Because the divergence
for each $\theta$ is strictly convex in $\omega_{Q^{*}}^{\prime}$,
so is the divergence in (\ref{eq:KL for stable}) for a fixed $\mu^{*}$.
Therefore, if $\omega^{\prime}$ and $\omega^{\prime\prime}$ both
minimize it, then $\omega_{Q^{*}}^{\prime}=\omega_{Q^{*}}^{\prime\prime}$.
By Proposition \ref{prop: long run beliefs}, any stable $\mu^{*}$
attaches probability one to that value of $\omega_{Q^{*}}$.

Notice also that by the same arguments as in Proposition \ref{prop: constant rage-1},
the marginal of $\mu_{t}$ on $\omega_{-Q^{*}}$ is equal to the marginal
of the prior. Denote $\bar{\theta}(\omega_{Q^{*}})$ to be $\bar{\theta}(\mu^{\omega_{Q^{*}}})$,
where $\mu^{\omega_{Q^{*}}}$ agrees with the prior over $\omega_{-Q^{*}}$
and attaches probability one to $\omega_{Q^{*}}$. By the implicit
function theorem and that all primitive functions are smooth, $\bar{\theta}(\omega_{Q^{*}})$
is a smooth function of $\omega_{Q^{*}}$ 

For any model $\left(q,K\right)$ and state $\omega^{*}$, define
$\bar{\theta}\left(\omega_{Q^{*}}^{*},q,K\right)$ so that 
\[
D\left(\int q\left(\cdot|\bar{\theta}\left(\omega_{Q^{*}}^{*},q,K\right),\omega\right)d\mu^{\omega_{Q^{*}}^{*}}||\int q\left(\cdot|\theta^{*},\omega\right)d\mu^{\omega_{Q^{*}}^{*}}\right)=K,
\]
i.e. $\bar{\theta}\left(\omega_{Q^{*}}^{*},q,K\right)$ is the cutoff
context when $\omega^{*}$ is true and the researcher's beliefs attach
probability $1$ to $\omega_{Q^{*}}^{*}$, given the model $\left(q,K\right)$.
For any model $\left(q,K\right)$ and $j\in Q$, define
\[
G_{j}(w,q,K)=\sum_{s}\int_{0}^{\bar{\theta}\left(\omega_{Q^{*}}^{*},q,K\right)}q\left(s|\theta,w\right)\frac{\frac{d}{dw_{j}}q\left(s|\theta^{*},w\right)}{q\left(s|\theta^{*},w\right)}p_{\theta}(\theta)d\theta
\]
 for $w\in\Omega$. Notice that $G_{j}\left(\omega^{*},q,K\right)$
is
\[
\frac{d}{d\omega_{j}}D_{KL}\left(\left(\int_{0}^{\bar{\theta}\left(\omega_{Q^{*}}^{*},q,K\right)}q\left(\cdot|\theta,\omega^{*}\right)p_{\theta}\left(\theta|\theta\leq\bar{\theta}\left(\omega_{Q^{*}}^{*},q,K\right)\right)d\theta\right)||q\left(\cdot|\theta^{*},\omega\right)\right)|_{\omega=\omega^{*}}
\]
multiplied by the probability that $\theta^{t}$ is less than $\bar{\theta}\left(\omega_{Q^{*}}^{*},q,K\right)$,
which does not affect whether the expression is equal to zero. Therefore,
if $\mu^{*}$ is a stable belief for $\omega^{*}$ attaching probability
$1$ to $\omega_{Q}^{*}$, then $G_{j}\left(\omega^{*},p_{S},K\right)=0$.

We want to show that $\left\{ \omega\in\Omega:G_{j}\left(\omega,q,K\right)=0\right\} $
has measure zero for ``most'' regular models $\left(q,K\right)$.
Since $\mu$ is a product measure that admits a density with support
$\Omega$ and $\Omega$ is compact and convex, $\Omega$ is a manifold
with boundary of dimension $n$, and we can work only with its interior
since its boundary has dimension $n-1$ \citep[p. 59,][]{guillemin1974differential}
and so zero measure. Towards that end, pick any regular model $\left(q,K\right)$
and its corresponding $Q^{*}$. Fix $k\notin Q^{*}$ and a $j\in Q$
so that for every $\omega$, $\frac{d}{d\omega_{j}}q(s|\omega,\theta^{*})\neq0$
for some $s$. Pick any $s^{*}\in S$ and define
\[
q\left(\zeta\right)\left(s^{\prime}|\theta,w\right)=q\left(s^{\prime}|\theta,w\right)+\sum_{s\neq s^{*}}\left(\mathbb{I}_{\{s\}}\left(s^{\prime}\right)-\mathbb{I}_{\{s^{*}\}}\left(s^{\prime}\right)\right)\omega_{k}\theta\zeta_{s}
\]
for every $\zeta\in B_{\varepsilon}(0)=Z\subset\mathbb{R}^{\#S-1}$
for a sufficiently small $\varepsilon>0$. When $\varepsilon$ is
small enough, $\left(q\left(\zeta\right),K^{\prime}\right)$ is a
regular model with the same $\theta^{*}$ and $Q^{*}$ as $q$ for
every $\zeta\in Z$ and $K^{\prime}>0$, and $q\left(\zeta\right)$
has a strictly convex divergence whenever $q$ does.\footnote{We can extend $Z$ to paramterize any family of polynomials with the
same $\theta^{*}$, $Q^{*}$, and maximum degree in a similar fashion
without changing arguments.}

Note that $q\left(\zeta\right)\left(\cdot|\theta^{*},w\right)=q\left(\cdot|\theta^{*},w\right)$
for all $w$. Then, $\frac{dG_{j}\left(w,q\left(\zeta\right),K\right)}{d\zeta_{s}}$
equals
\begin{align*}
 & w_{k}\left(\frac{\frac{d}{dw_{j}}q\left(s|\theta^{*},w\right)}{q\left(s|\theta^{*},w\right)}-\frac{\frac{d}{dw_{j}}q\left(s^{*}|\theta^{*},w\right)}{q\left(s^{*}|\theta^{*},w\right)}\right)\int_{0}^{\bar{\theta}\left(w_{Q^{*}},q(\zeta),K\right)}\theta p_{\theta}(\theta)d\theta\\
 & +\frac{d\bar{\theta}\left(w_{Q^{*}},q\left(\zeta\right),K\right)}{d\zeta_{s}}\sum_{s}q\left(\zeta\right)\left(s|\bar{\theta}\left(w_{Q^{*}},q(\zeta),K\right),w\right)\frac{\frac{d}{dw_{j}}q\left(s|\theta^{*},w\right)}{q\left(s|\theta^{*},w\right)}p_{\theta}\left(\bar{\theta}\left(w_{Q^{*}},q(\zeta),K\right)\right).
\end{align*}
and $\frac{dG_{j}\left(w,q\left(\zeta\right),K\right)}{dK}$ equals
\[
\frac{d\bar{\theta}\left(w_{Q^{*}},q(\zeta),K\right)}{dK}\sum_{s}q\left(\zeta\right)\left(s|\bar{\theta}\left(w_{Q^{*}},q(\zeta),K\right),w\right)\frac{\frac{d}{dw_{j}}q\left(s|\theta^{*},w\right)}{q\left(s|\theta^{*},w\right)}p_{\theta}\left(\bar{\theta}\left(w_{Q^{*}},q(\zeta),K\right)\right).
\]
By our definition of $j\in Q$, for every $w$ there exist $s^{\prime},s^{\prime\prime}\in S$
so that 
\[
\frac{d}{dw_{j}}q\left(s^{\prime}|\theta^{*},w\right)>0>\frac{d}{dw_{j}}q\left(s^{\prime\prime}|\theta^{*},w\right)
\]
so for some $s^{**}\in\left\{ s^{\prime},s^{\prime\prime}\right\} $,
\[
\frac{\frac{d}{dw_{j}}q\left(s^{**}|\theta^{*},w\right)}{q\left(s|\theta^{*},w\right)}\neq\frac{\frac{d}{dw_{j}}q\left(s^{*}|\theta^{*},w\right)}{q\left(s^{*}|\theta^{*},w\right)}.
\]
When $w_{k}\neq0$, either 
\[
\sum_{s}q\left(\zeta\right)\left(s|\bar{\theta}\left(w_{Q^{*}}\right),w\right)\frac{\frac{d}{dw_{j}}q\left(s|\theta^{*},w\right)}{q\left(s|\theta^{*},w\right)}p_{\theta}(\bar{\theta}(\omega_{Q^{*}}^{*}))=0
\]
 and then $\frac{dG_{j}\left(w,q\left(\zeta\right),K\right)}{d\zeta_{s^{**}}}\neq0$,
or it does not equal zero, in which case $\frac{dG_{j}\left(w,q\left(\zeta\right),K\right)}{dK}\neq0$
(since $\frac{d\bar{\theta}\left(w_{Q^{*}},q(\zeta),K\right)}{dK}\neq0$).
Therefore, $G_{j}$ viewed as a function from $\Omega\setminus\left\{ \omega:\omega_{k}=0\right\} \times Z\times\mathbb{R}_{++}$
to $\mathbb{R}$ is transversal to $\left\{ 0\right\} $. By the Transversality
Theorem \citep[p 68,][]{guillemin1974differential},
\[
\left\{ \omega\in\Omega:G_{j}\left(\omega,q\left(\zeta\right),K^{\prime}\right)=0\right\} 
\]
is a dimension $n-1$ subset of $\Omega$ for almost all $\left(\zeta,K^{\prime}\right)$
in $Z\times\mathbb{R_{++}}$. That is, for any regular $(q,K)$, there
are many $q\left(\zeta\right)$ arbitrarily close to $q$ and cutoffs
$K'$ arbitrarily close to $K$ for which the set of fixed parameters
$\omega^{*}$ that have $\omega^{*}$ as a minimizer of (\ref{eq:KL for stable})
has measure zero. By Proposition \ref{prop: long run beliefs} and
the first part of this proposition, no stable beliefs for any of those
parameters attach positive probability to $\omega_{Q}^{*}$.

\subsection{Proof of Proposition \ref{prop: multiple steady states}}

Let $\bar{\mu}_{1}^{t}$ denote the mean belief about $\omega_{1}$
according to the belief $\mu_{t}$, and $\bar{\mu}_{2}$ denote the
time-invariant mean belief about $\omega_{2}$. Denote $q^{t}=(1-\theta^{t})\bar{\mu}_{1}^{t}+\theta^{t}\bar{\mu}_{2}$.
The KL divergence that determines whether research is conducted in
period $t$ is

\[
D_{KL}\left(p_{S}(\cdot|\theta^{t},h^{t})||p_{S}(\cdot|\theta^{*},h^{t})\right)=q^{t}\ln\frac{q^{t}}{\bar{\mu}_{1}^{t}}+\left(1-q^{t}\right)\ln\frac{1-q^{t}}{1-\bar{\mu}_{1}^{t}},
\]
a function of only $\theta^{t}$ and $\bar{\mu}_{1}^{t}$. Moreover,
$\varTheta^{R}\left(\mu_{t}\right)$ is an interval $\left[0,\bar{\theta}\left(\bar{\mu}_{1}^{t}\right)\right]$
.

For $\bar{\theta}\in\left[0,1\right]$, define the expected frequency
with which $s^{t}=1$ given $\theta\leq\bar{\theta}$ and $\omega^{*}$
as
\[
s\left(\bar{\theta},\omega^{*}\right)=E\left[\theta|\theta\leq\bar{\theta}\right]\cdot(\omega_{2}^{*}-\omega_{1}^{*})+\omega_{1}^{*}.
\]
Given $a_{1}>\frac{1}{2}$ and $\zeta>0$, we show that for every
$K$, there exists $\delta>0$ so that $\hat{w}_{1}=\frac{1}{2}\left(\omega_{2}^{*}+\omega_{1}^{*}\right)\in\left(\frac{1-\zeta}{2},\frac{1+\zeta}{2}\right)$
is a solution to (\ref{eq: steady-state equation}) when $\omega^{*}\in B_{\delta}\left(a_{1},1-a_{1}\right)$.
To see why, note that when $\bar{\mu}_{1}^{t}=\frac{1}{2}$, we have
$q^{t}=\bar{\mu}_{1}^{t}$ for every $\theta^{t}$. This means that
$D_{KL}\left(p_{S}(\cdot|\theta^{t},h^{t})||p_{S}(\cdot|\theta^{*},h^{t})\right)=0$
for all $\theta^{t}\in[0,1]$, hence $\bar{\theta}\left(\frac{1}{2}\right)=1$.
By continuity, there exists $\delta^{*}>0$ so that $\bar{\theta}\left(x\right)=1$
for all $x\in\left(\frac{1-\delta^{*}}{2},\frac{1+\delta^{*}}{2}\right)$.
Notice that when $\omega^{*}\in B_{\delta}\left(a_{1},1-a_{1}\right)$,
\[
s\left(1,\omega^{*}\right)\in\left[\frac{1}{2}-\delta,\frac{1}{2}+\delta\right].
\]
Take $\delta=\min\left\{ \delta^{*},\zeta\right\} $, and then $\hat{w}_{1}\left(\omega^{*}\right)=s\left(1,\omega^{*}\right)\in\left(\frac{1-\zeta}{2},\frac{1+\zeta}{2}\right)$
satisfies Equation (\ref{eq: steady-state equation}) for all $\omega^{*}\in E$.

For small enough $K$ and $\zeta$, a second steady state exists for
the same $\omega^{*}\in B_{\delta}\left(a_{1},a_{2}\right)$ as above.
Notice that when $\zeta<\min\left\{ \frac{1}{3}\left(2a_{1}-1\right),1-\varepsilon-a_{1}\right\} $,
\[
1-\varepsilon>a_{1}+\zeta>a_{1}-\zeta>\frac{1+\zeta}{2}>\frac{1}{2}\left(\omega_{2}^{*}+\omega_{1}^{*}\right),
\]
$\frac{d}{d\bar{\theta}}s\left(\bar{\theta},\omega^{*}\right)<0$,
$s(0,\omega^{*})=\omega_{1}^{*}$, and $s(1,\omega^{*})=\frac{1}{2}\omega_{2}^{*}+\frac{1}{2}\omega_{1}^{*}$.
Pick $\theta^{\prime}>0$ so that $s\left(\theta,\omega^{*}\right)>\frac{1+\zeta}{2}$
for all $\theta<\theta^{\prime}$ and $\omega^{*}$ as above. For
every small $\bar{\theta}\in\left(0,\theta^{\prime}\right)$, define
\begin{align*}
K(\bar{\theta})\equiv & \left[\left(1-\bar{\theta}\right)\left(a_{1}-\zeta\right)+\bar{\theta}\frac{1}{2}\right]\ln\frac{\left(1-\bar{\theta}\right)\left(a_{1}-\zeta\right)+\bar{\theta}\frac{1}{2}}{a_{1}+\zeta}\\
 & +\left[\left(1-\bar{\theta}\right)\left(1-a_{1}+\zeta\right)+\bar{\theta}\frac{1}{2}\right]\ln\frac{\left(1-\bar{\theta}\right)\left(1-a_{1}-\zeta\right)+\bar{\theta}\frac{1}{2}}{1-a_{1}+\zeta}
\end{align*}
Because $D_{KL}\left(p_{S}\left(\cdot|\bar{\theta},h_{t}\right)||p_{S}\left(\cdot|\theta^{*},h_{t}\right)\right)$
equals
\[
\left[\left(1-\bar{\theta}\right)\bar{\mu}_{1}^{t}+\bar{\theta}\frac{1}{2}\right]\ln\frac{\bar{\theta}\frac{1}{2}+\left(1-\bar{\theta}\right)\bar{\mu}_{1}}{\bar{\mu}_{1}}+\left[\left(1-\bar{\theta}\right)\left(1-\bar{\mu}_{1}^{t}\right)+\bar{\theta}\frac{1}{2}\right]\ln\frac{1-\left(\bar{\theta}\frac{1}{2}+\left(1-\bar{\theta}\right)\bar{\mu}_{1}\right)}{1-\bar{\mu}_{1}},
\]
$K(\bar{\theta})$ is weakly below $D_{KL}\left(p_{S}\left(\cdot|\bar{\theta},h_{t}\right)||p_{S}\left(\cdot|\theta^{*},h_{t}\right)\right)$
for all histories so that $\bar{\mu}_{1}^{t}\in\left[a_{1}-\zeta,a_{1}+\zeta\right]$.
Therefore, when $K<K\left(\bar{\theta}\right)$ and $\bar{\mu}_{1}^{t}\in\left[a_{1}-\zeta,a_{1}+\zeta\right]$,
$\bar{\theta}\left(\bar{\mu}_{1}^{t}\right)\leq\bar{\theta}$. This
implies
\[
a_{1}-\zeta\leq s(\bar{\theta},w^{*})\leq E\left[\theta|\theta\leq\bar{\theta}\left(\bar{\mu}_{1}^{t}\right)\right]\cdot(w_{2}^{*}-w_{1}^{*})+w_{1}^{*}\leq1-\varepsilon
\]
By the intermediate value theorem, for any $K\in\left(0,K\left(\bar{\theta}\right)\right)$,
there exists a $\hat{\hat{w}}_{1}\in\left[a_{1}-\zeta,1-\varepsilon\right]$
solving Equation (\ref{eq: steady-state equation}). Therefore, both
$\hat{w}_{1}$ and $\hat{\hat{w}}_{1}$ both solve Equation (\ref{eq: steady-state equation})
for $\omega^{*}$.

Moreover, the two solutions are attractors of the dynamic process.
For a mean belief $x$ that is sufficiently close to either solution,
there are fewer (more) ``success'' realizations $s=1$ than expected
when $x$ is above (below) the fixed point. For $\omega_{1}=\hat{\omega}_{1}\left(\theta_{2}(K)\right)$,
this follows from there being too few successes at $1-\varepsilon$
and too many successes at $\frac{1}{2}$ . For $\omega_{1}=\frac{1}{2}$,
this follows because $\bar{\theta}\left(x\right)=1$ for all $x$
sufficiently close to $\frac{1}{2}$, and so the number of successes
is locally constant in $x$. Consequently, when a belief is near one
of these two fixed points, it tends to drift toward it. $\oblong$

\subsection{Proof of Proposition \ref{prop:calibration}}

Under the identifying assumption that $\omega_{-i}=m_{-i}^{t}$, beliefs
evolve so that $m_{-i}^{t+1}=m_{-i}^{t}$ and 
\begin{align*}
m_{i}^{t+1} & =m_{i}^{t}+\frac{(\sigma_{i}^{t})^{2}}{(\sigma_{i}^{t})^{2}+1}\left(s^{t}-m_{1}^{t}-m_{2}^{t}\right)\\
 & =\frac{1}{(\sigma_{i}^{t})^{2}+1}m_{i}^{t}+\frac{(\sigma_{i}^{t})^{2}}{(\sigma_{i}^{t})^{2}+1}\left(s^{t}-m_{-i}^{t}\right)
\end{align*}
by the usual formula for updating a Normal distribution. Suppose that
$\left(\sigma_{i}^{0}\right)^{2}=v$ for $i=1,2$, and that $K$ is
large enough that research is conducted at $t=1$. W.l.o.g, the researcher
updates her beliefs over $\omega_{1}$ ($\omega_{2}$) in odd (even)
periods.

Break the time periods into blocks: block $1$ corresponds to $t=1,2$;
block $2$ corresponds to $t=3,4$; etc. Let $s(\tau,k)$ denote the
$s$ realization in part $k$ of block $\tau$. Then, the variance
after block $\tau$ is
\[
\sigma_{1}^{2}(\tau)=\sigma_{2}^{2}(\tau)=\frac{v}{1+\tau v}
\]
Denote
\[
\alpha_{\tau}=\frac{1+\tau v}{1+(1+\tau)v}
\]
The updated means $m_{1}(\tau+1)$ and $m_{2}(\tau+1)$ at the end
of block $\tau+1$ are given by
\begin{equation}
m_{1}(\tau+1)=\alpha_{\tau}m_{1}(\tau)+(1-\alpha_{\tau})(s(\tau+1,1)-m_{2}(\tau))\label{eq:mean evolution}
\end{equation}
and

\begin{align*}
m_{2}(\tau+1) & =\alpha_{\tau}m_{2}(\tau)+(1-\alpha_{\tau})(s(\tau+1,2)-m_{1}(\tau+1))\\
 & =\alpha_{\tau}m_{2}(\tau)+(1-\alpha_{\tau})(s(\tau+1,2)-m_{1}(\tau))-(1-\alpha_{\tau})^{2}(s(\tau+1,1)-m_{1}(\tau))-m_{2}(\tau))\\
 & =(\alpha_{\tau}+(1-\alpha_{\tau})^{2})m_{2}(\tau)+(1-\alpha_{\tau})s(\tau+1,2)-(1-\alpha_{\tau})^{2}s(\tau+1,1)-(1-\alpha_{\tau})\alpha_{\tau}m_{1}(\tau).
\end{align*}
Add up the two equations for $m_{i}(\tau+1)$ and denote
\begin{align*}
x(\tau+1) & =m_{1}(\tau+1)+m_{2}(\tau+1)\\
 & =\alpha_{\tau}^{2}x(\tau)+(1-\alpha_{\tau}^{2})\left[\frac{\alpha_{\tau}}{1+\alpha_{\tau}}s(\tau+1,1)+\frac{1}{1+\alpha_{\tau}}s(\tau+1,2)\right].
\end{align*}
We first consider the distribution of $x(\tau+1)$, then that of $m_{i}(\tau)$.

Since $x(0)$ is a given constant, we can write 
\[
x(1)=\beta_{0}^{1}x(0)+\beta_{1}^{1}s_{1}+\beta_{2}^{1}s_{2}
\]
 with $\beta_{1}^{1},\beta_{2}^{1}\leq1-\alpha_{0}$. For $\tau\geq1$,
suppose that 
\[
x(\tau)=\beta_{0}^{\tau}x(0)+\beta_{1}^{\tau}s_{1}+\dots+\beta_{2\tau}^{\tau}s_{2\tau}
\]
 with $\beta_{j}^{\tau}\leq1-\alpha_{\tau-1}$ for each $j>0$. Then,
\[
x(\tau+1)=\alpha_{\tau}^{2}(\beta_{0}^{\tau}x(0)+\beta_{1}^{\tau}s_{1}+\dots+\beta_{2\tau}^{\tau}s_{2\tau})+\alpha_{\tau}(1-\alpha_{\tau})s_{2\tau+1}+(1-\alpha_{\tau})s_{2\tau+2}.
\]
For all $0<j\leq2\tau$, when we let
\[
\beta_{j}^{\tau+1}\equiv\alpha_{\tau}^{2}\beta_{j}^{\tau}\leq\alpha_{\tau}\beta_{j}^{\tau}\leq\alpha_{\tau}\left(1-\alpha_{\tau-1}\right)=\frac{1+\tau v}{1+(1+\tau)v}\cdot\frac{1}{1+\tau v}=1-\alpha_{\tau},
\]
it follows that 
\begin{equation}
x(\tau+1)=\beta_{0}^{\tau+1}x(0)+\beta_{1}^{\tau+1}s_{1}+\dots+\beta{}_{2\tau+2}^{\tau+1}s_{2\tau+2}\label{eq:x(u)}
\end{equation}
with $\beta_{j}^{\tau+1}\leq1-\alpha_{\tau}$ for all $j>0$.

By the above, $x(\tau)|\omega\sim N\left(m_{\tau},v_{\tau}\right)$
with 
\[
v_{\tau}\leq\sum_{j=1}^{2\tau}\left(\beta_{j}^{\tau}\right)^{2}\leq2\tau\left[1-\alpha_{\tau-1}\right]^{2}=\frac{2\tau v^{2}}{\left(1+\tau v\right)^{2}}
\]
 for all $\tau>1$. This upper bound tends to zero as $\tau\rightarrow\infty$.
Finally, notice that 
\[
\beta_{0}^{\tau+1}=\prod_{j=0}^{\tau}\alpha_{\tau}^{2}=\prod_{j=0}^{\tau}\frac{(1+jv)^{2}}{\left(1+(1+j)v\right)^{2}}=\left(\frac{1}{1+\left(\tau+1\right)v}\right)^{2}\rightarrow0
\]
as $\tau\rightarrow\infty$, and that $\beta_{0}^{\tau+1}+\beta_{1}^{\tau+1}+\dots+\beta{}_{2\tau+2}^{\tau+1}=1$.
Therefore, in the $\tau\rightarrow\infty$ limit, $x(\tau+1)$ in
(\ref{eq:x(u)}) is a convex combination of $s$ realizations. Hence,
$x(\tau+1)\rightarrow\mathbb{E}[s_{i}|\omega]=\omega_{1}+\omega_{2}$.

We now turn to beliefs about $\omega_{i}$. Using recursive substitutions
of Equation (\ref{eq:mean evolution}), we show by induction that
\begin{equation}
m_{i}(\tau)=k_{0}^{i,\tau}+(-1)^{i}\sum_{j=1}^{\tau}k_{j,2}^{i,\tau}s(j,2)+(-1)^{i+1}\sum_{j=1}^{\tau}k_{j,1}^{i,\tau}s(j,1)\label{eq:mit in terms of k}
\end{equation}
 for some $k_{j,h}^{i,\tau}\in[(1-\alpha_{j-1})\alpha_{j-1},1-\alpha_{j-1}]$
for $1\leq j<\tau$, $k_{\tau,2}^{1,\tau}=0$, $k_{\tau,1}^{2,\tau}=\alpha_{\tau-1}\left(1-\alpha_{\tau-1}\right)$,
and $k_{\tau,1}^{1,\tau}=k_{\tau,2}^{2,\tau}=1-\alpha_{\tau-1}$.
In particular, $m_{1}(\tau)$ is increasing in odd signals and decreasing
in even signals, and vice versa for $m_{2}(\tau)$. If true, then
non-vanishing weight gets placed on every signal.

Equation (\ref{eq:mit in terms of k}) holds with weights in appropriate
bounds for $m_{1}(1)$ since 
\[
m_{1}(1)=(1-\alpha_{0})s_{1}+k_{0}^{1,1}
\]
with $k_{0}^{1,1}=\alpha_{0}m_{1}(0)$, $k_{1,1}^{1,1}=(1-\alpha_{0})$
and $k_{1,2}^{1,1}=0$. Also for $m_{2}(1)$ since
\[
m_{2}(1)=(1-\alpha_{0})s_{2}-\alpha_{0}(1-\alpha_{0})s_{1}+k_{0}^{2,1}
\]
with $k_{0}^{2,1}=\alpha_{0}m_{2}(0)$, $k_{1,1}^{2,1}=\alpha_{0}(1-\alpha_{0})$
and $k_{1,2}^{2,1}=(1-\alpha_{0})$.

Assume that there exist weights $k_{j,h}^{i,\tau}$ as claimed so
that equation (\ref{eq:mit in terms of k}) holds for $\tau$ and
$i=1,2$. Substituting the inductive hypothesis into equation (\ref{eq:mean evolution}),
\begin{align*}
m_{1}(\tau+1)= & \alpha_{\tau}m_{1}(\tau)+(1-\alpha_{\tau})s(\tau+1,1)-(1-\alpha_{\tau})m_{2}(\tau)\\
= & \sum_{j=1}^{\tau}[\alpha_{\tau}k_{j,1}^{1,\tau}+(1-\alpha_{\tau})k_{j,1}^{2,\tau}]s(j,1)+(1-\alpha_{\tau})s(\tau+1,1)\\
 & \qquad-\sum_{j=1}^{\tau}[\alpha_{\tau}k_{j,2}^{1,\tau}+(1-\alpha_{\tau})k_{j,2}^{2,\tau}]s(j,2)+[\alpha_{\tau}k_{0}^{1,\tau}-(1-\alpha_{\tau})k_{0}^{2,\tau}].
\end{align*}
Equation (\ref{eq:mit in terms of k}) holds for $\tau+1$ and $i=1$
when we let $k_{0}^{1,\tau+1}=\alpha_{\tau}k_{0}^{1,\tau}-(1-\alpha_{\tau})k_{0}^{2,\tau}$,
$k_{\tau+1,1}^{1,\tau+1}=(1-\alpha_{\tau})$, $k_{\tau+1,2}^{1,\tau+1}=0$,
and $k_{j,h}^{1,\tau+1}=\alpha_{\tau}k_{j,h}^{1,\tau}+(1-\alpha_{\tau})k_{j,h}^{2,\tau}$
for $h=1,2$ and $j\leq\tau$. These are clearly within the bounds.
Similarly,
\begin{align*}
m_{2}(\tau+1)= & \sum_{j=1}^{\tau}[\alpha_{\tau}k_{j,2}^{2,\tau}+(1-\alpha_{\tau})k_{j,2}^{1,\tau}]s(j,2)+\left(1-\alpha_{\tau}\right)s(\tau+1,2)-\alpha_{\tau}(1-\alpha_{\tau})s(\tau+1,1)\\
 & \qquad-\sum_{j=1}^{\tau}[\alpha_{\tau}k_{j,1}^{2,\tau}+(1-\alpha_{\tau})k_{j,1}^{1,\tau}]s(j,1)+[\alpha_{\tau}k_{0}^{2,\tau}+(1-\alpha_{\tau})k_{0}^{1,\tau}]
\end{align*}
so $k_{j,h}^{2,\tau+1}$ can be defined in a similar way so that equation
(\ref{eq:mit in terms of k}) holds for $\tau+1$ and $i=2$. Inductive
arguments extend the formula to all $\tau$.

Now, observe that $m_{i}(\tau)$ is a normally distributed random
variable. Conditional on $\omega_{1}+\omega_{2}$, its variance is
bounded from below by, say, $(k_{1,1}^{i,\tau+1})^{2}\geq((1-\alpha_{1})\alpha_{1})^{2}>0$.
It is bounded from above by 
\[
\sum_{j=1}^{\tau-1}[(k_{j,1}^{i,\tau})^{2}+(k_{j,2}^{i,\tau})^{2}]\leq2\sum_{j=1}^{\infty}\left(1-\alpha_{j}\right)^{2}=2\sum_{j=1}^{\infty}\left(\frac{v}{1+jv}\right)^{2}.
\]
This sum converges by the integral rule. $\oblong$

\subsection{Proof of Proposition \ref{prop: Heckman}}

For almost every history $h^{t}$, $\mu\left(h^{t}\right)$ is normally
distributed with variables independent. Let $\eta$ denote any such
beliefs with $\eta_{i}$ the marginal on the $i$-th dimension. Slightly
abusing notation,\footnote{Namely, by the ``conditioning'' on $\eta$. The meaning is that
the distribution $p_{S,U}$ is induced by the distribution $\eta$
over $\omega$.}
\[
S(\eta,\theta)=D_{KL}\left(p_{S,U}(\cdot|\eta,\theta)||p_{S,U}(\cdot|\eta_{1},\eta_{3},\omega_{2}^{*}=0,\theta)\right)
\]
 and 
\[
R\left(\eta,\theta\right)=D_{KL}\left(p_{S,U}(\cdot|\eta,\theta)||p_{S,U}(\cdot|\eta,\theta^{*}=0)\right).
\]
 Denote $g(x)=x-\ln x-1$, noting that $g^{\prime}(x)>0$ when $x>1$
and that $g(1)=0$, and
\[
h(x,y)=x\ln\left(\frac{x}{y}\right)+(1-x)\ln\left(\frac{1-x}{1-y}\right).
\]
Then,
\begin{align*}
S(\eta,\theta)= & \frac{1}{4}\left[g\left(1+\frac{\sigma_{2}^{2}}{\sigma_{1}^{2}+\lambda_{1}^{2}\theta^{2}\sigma_{3}^{2}}\right)+\frac{m_{2}^{2}}{\sigma_{1}^{2}+\lambda_{1}^{2}\theta^{2}\sigma_{3}^{2}}\right]\\
R(\eta,\theta)= & \frac{1}{4}\left[g\left(1+\frac{\lambda_{1}^{2}\theta^{2}\sigma_{3}^{2}}{\sigma_{2}^{2}+\sigma_{1}^{2}}\right)+\frac{\lambda_{1}^{2}\theta^{2}m_{3}^{2}}{\sigma_{2}^{2}+\sigma_{1}^{2}}+g\left(1+\frac{\lambda_{0}^{2}\theta^{2}\sigma_{3}^{2}}{\sigma_{1}^{2}}\right)+\frac{\lambda_{0}^{2}\theta^{2}m_{3}^{2}}{\sigma_{1}^{2}}\right]+D_{S_{1}|S_{2},U}(\theta)
\end{align*}
where 
\[
\lambda_{i}=\mathbb{E}\left[u|s_{1}=1,s_{2}=i\right]=\frac{\phi(-i)}{1-\Phi(-i)},
\]
and
\[
D_{S_{1}|S_{2},U}(\theta)=\int\frac{1}{2}\phi(u)\left(h\left(\theta\Phi(-1-u)+(1-\theta)\Phi(-1),\Phi(-1)\right)+h\left(\theta\Phi(-u)+(1-\theta)\frac{1}{2},\frac{1}{2}\right)\right)du
\]
is the expected KL divergence of $p_{S_{1}}(\cdot|s_{2},u,\theta)$
from $p_{S_{1}}(\cdot|s_{2},u,\theta^{*}=0)$. This follows from the
formula for KL divergence of two normal distributions, and from the
observation that $D_{KL}\left(p_{S,U}(\cdot|\theta)||p_{S,U}(\cdot|\theta^{*}=0)\right)$
equals
\[
\sum_{s_{2}}p(s_{2})\int\left[D_{KL}\left(p_{S_{3}}(\cdot|\theta,s_{2},u)||p_{S_{3}}(\cdot|s_{2},\theta^{*}=0,u)\right)+D_{KL}\left(p_{S_{1}}(\cdot|\theta,s_{2},u)||p_{S_{1}}(\cdot|s_{2},\theta^{*}=0,u)\right)\right]d\Phi(u).
\]

Clearly, $S$ decreases in $\theta$, $R$ increases in $\theta$,
$R(\eta,0)=0$, and $S(\eta,0)>0$. Therefore, there is an interval
$[0,x]$ with $0<x$ such that $R(\eta,\theta)\geq S(\eta,\theta)$
if and only if $\theta\in[0,x]$. Similarly, there is an interval
$[0,y]$ with $y>0$ such that $R(\eta,\theta)\leq K$ if and only
if $\theta\in[0,y]$. Finally, there is an interval $(z,1]$ (with
$z$ possibly equal to $1$) such that $S(\eta,\theta)<K$ if and
only if $\theta\in(z,1]$. Then, $\left[0,\bar{\theta}^{RD}\left(\eta\right)\right]=[0,x]\cap[0,y]=\left[0,\min\left\{ x,y\right\} \right]$,
and $\left(\bar{\theta}^{S}\left(\eta\right),1\right]=(x,1]\cap(z,1]=\left(\max\left\{ x,z\right\} ,1\right]$.
In the former interval, $\theta^{*}=0$ induces a lower KL divergence
than does $\omega_{2}^{*}=0$, and the divergence is below $K$. In
the latter interval, $\omega_{2}^{*}=0$ induces a lower KL divergence
than does $\theta^{*}=0$, and the divergence is below $K$. If $K$
is sufficiently large, then $z=0$ and $y=1$, so the two intervals
are adjacent.

Notice that $S$ strictly increases in $m_{2}^{2}$, while $R$ is
constant in it. Therefore, an increase in $m_{2}^{2}$ leads to an
increase in $\bar{\theta}^{RD}\left(\eta\right)$ (weakly) and $\bar{\theta}^{S}\left(\eta\right)$
(strictly). Also, $R$ strictly increases in $m_{3}^{2}$, while $S$
is constant in it. Therefore, an increase in $m_{3}^{2}$ leads to
a decrease in $\bar{\theta}^{RD}\left(\eta\right)$ (strictly) and
$\bar{\theta}^{S}\left(\eta\right)$ (weakly). Finally, $S$ strictly
increases in $\sigma_{2}^{2}$, and $R$ strictly decreases in it.
Therefore, an increase in $\sigma_{2}^{2}$ leads to a (strict) decrease
in both $\bar{\theta}^{RD}\left(\eta\right)$ and $\bar{\theta}^{S}\left(\eta\right)$.
$\oblong$

\bibliographystyle{plainnat}
\bibliography{assumption}

\end{document}